%% file: arxiv.tex
\documentclass[a4paper,fleqn,usenatbib]{mnras}
\usepackage[T1]{fontenc}
\usepackage{ae,aecompl}
\usepackage{graphicx}
\usepackage{amsmath}
\usepackage{amssymb}
\usepackage{natbib}
\usepackage[utf8]{inputenc}
\usepackage{epstopdf}

\title[UV-bright, overluminous field stars]{A kinematically unbiased, all-sky search for nearby, young, low-mass stars}

\author[A. S. Binks et al.]{
Alexander S. Binks,$^{1,2}$\thanks{E-mail: a.s.binks1@keele.ac.uk}
Matthieu Chalifour,$^{3,4}$
Joel H. Kastner,$^{3}$
\newauthor{
David Rodriguez,$^{5}$
Simon J. Murphy,$^{6}$
David A. Principe,$^{7}$
Kristina Punzi,$^{8}$}
\newauthor{
Germano G. Sacco$^{9}$
and Jes\'us Hern\'andez$^{10}$
}
\\
$^{1}$Astrophysics Group, School of Chemistry and Physics, Keele University, UK \\
$^{2}$Instituto de Radioastronom\'ia y Astrof\'isica, Universidad Nacional Aut\'onoma de M\'exico (UNAM), Morelia, M\'exico \\
$^{3}$Center for Imaging Science and Laboratory for Multiwavelength Astrophysics, Rochester Institute of Technology, Rochester, NY, USA \\ 
$^{4}$Department of Physics \& Astronomy, Swarthmore College,  Swarthmore, PA, USA \\
$^{5}$Space Telescope Science Institute, Baltimore, MD, USA \\
$^{6}$School of Science, The University of New South Wales, Canberra, ACT 2600, Australia \\
$^{7}$MIT Kavli Institute for Astrophysics and Space Research, 70 Vassar St, Cambridge, MA 02109, USA \\
$^{8}$Astronomy Department, Wellesley College, Wellesley, MA, USA \\
$^{9}$Arcetri Observatory, Florence, Italy \\
$^{10}$Instituto de Astronom\'ia, UNAM, Unidad Acad\'emica en Ensenada, Ensenada 22860, M\'exico}
\date{Accepted XXX. Received YYY; in original form ZZZ}
\pubyear{2019}

\graphicspath{{/home/asbinks/chalifour2018/LatEx/figs/}}

\begin{document}
\label{firstpage}
\pagerange{\pageref{firstpage}--\pageref{lastpage}}
\maketitle

\begin{abstract}
The past two decades have seen dramatic progress in our knowledge of the population of young stars of age $< 200\,$Myr that lie within $150\,$pc of the Sun. These nearby, young stars, most of which are found in loose, comoving groups, provide the opportunity to explore (among many other things) the dissolution of stellar clusters and their diffusion into the field star population. Here, we exploit the combination of astrometric and photometric data from {\it Gaia} and photometric data from GALEX (UV) and 2MASS (near-IR) in an attempt to identify additional nearby, young, late-type stars. Specifically, we present a sample of 146 GALEX UV-selected late-type (predominantly K-type) field stars with {\it Gaia}-based distances $< 125\,$pc (based on {\it Gaia} Data Release 1) that have isochronal ages $< 80\,$Myr even if equal-components binaries. We investigate the spectroscopic and kinematic properties of this sample. Despite their young isochronal ages, only $\sim 10$ per cent) of stars among this sample can be confidently associated with established nearby, young moving groups (MGs). These candidate MG members include 5 stars newly identified in this study. The vast majority of our sample of 146 nearby young star candidates have anomalous kinematics relative to the known MGs. These stars may hence represent a previously unrecognised population of young stars that has recently mixed into the older field star population. We discuss the implications and caveats of such a hypothesis---including the intriguing fact that, in addition to their non-young-star-like kinematics, the majority of the UV-selected, isochronally young field stars within $50\,$pc appear surprisingly X-ray faint.
\end{abstract}


\section{Introduction}\label{sec:intro}

The identification and study of stars of age $< 200\,$Myr within $\sim 100\,$pc of the Sun provides crucial insight into pre-main sequence (pre-MS) stellar evolution and the formative stages of planetary systems \citep{2016a_Kastner}. Such young, nearby stars provide excellent samples for direct-imaging campaigns to observe exoplanets, circumstellar discs and brown dwarfs \citep[e.g.,][]{2004a_Kalas, 2009a_Lagrange, 2015a_Bowler, 2015a_MacGregor, 2015a_Chauvin}, act as direct observational test-beds for early stellar evolution \citep[e.g.,][]{2004a_Zuckerman,2015a_Bell}, and provide key evidence for the physical origins of young stars in the Solar neighbourhood and how they eventually disperse into the field population \citep[e.g.,][]{2018a_Wright}. 

The majority of these nearby, young stars can be placed in kinematically coherent ensembles known as nearby young moving groups (herein, MGs). To date, at least a dozen, and perhaps as many as two dozen, MGs have been identified \citep{2016a_Mamajek,2018a_Gagne}. Their members are amongst the youngest stars known in the Solar neighbourhood. These MGs are likely to have formed from members of more distant, dense star forming regions \citep[][]{2008a_Fernandez}, whose individual velocities are greater than the internal velocity dispersion, and subsequently escape in small groups. Since the ages for members of MGs can be resolved down to a few Myr, and MG member stars are approximately coeval \citep[age spreads generally $< 5\,$Myr;][]{2015a_Bell}, any age determination methods for a star in a MG can reasonably be applied to any other star in the group; furthermore, age determinations from diverse methods can create a tight age constraint for the MG \citep[e.g.,][]{2014a_Mamajek}. Recent work suggests that MGs share similar chemical abundances \citep{2013a_DeSilva,2013a_Barenfeld}, which provides evidence for their common origins and hints at the compositions of the molecular clouds from which they were born.

Over the past two decades, the identification of candidate nearby, young, late-type stars and (hence) MG members among the field-star population has proceeded via some combination of their signature luminous chromospheric (UV) and coronal (X-ray) emission, which result from strong surface magnetic fields \citep[e.g.,][and references therein]{1997a_Kastner,2013a_Rodriguez}, and their common Galactic ($UVW$) space motions \citep[e.g.,][]{2004a_Zuckerman,2008a_Torres,2013a_Malo}. Follow-up spectroscopy then further constrains stellar ages via determinations of Li absorption line strengths, rotation rates, and optical activity indicators (such as H$\alpha$ and the Ca {\sc ii} H+K lines and infrared triplet), so as to assess the viability of candidate MG stars or of proposed new MGs \citep[see discussions in][]{2004a_Zuckerman, 2015a_Binks}. 

Because MGs are sparse, and have spatial extents up to a few tens of pc \citep{2018a_Gagne}, their members can span wide areas on the celestial sphere. Hence, MG candidate selection has become increasingly reliant on kinematic and distance information, in addition to celestial coordinates and photometry. The fact that so many field stars, even within the nearest 100\,pc, were missing parallaxes, precise proper motions, and/or radial velocity measurements presented a major difficulty for previous searches for MG candidates and tests of their membership status \citep[e.g.,][]{2014a_Malo}. With the sudden availability of such data, in the form of the first two data releases from the {\it Gaia} space astrometry mission \citep[Data Releases 1 and 2, hereafter DR1 and DR2;][]{2016a_Gaia_Collaboration,2018a_Gaia_Collaboration}, the study of nearby, young stars and MGs can now make major strides. This potential motivated the recent study, described in \citet[][]{2017a_Kastner}, in which we evaluated the distances and ages of all 19 nearby young star candidates from the sample of $\sim 2000$ stars compiled by the GALEX \citep[UV,][]{2012a_Bianchi} Nearby Young Star Survey \citep[GALNYSS;][]{2013a_Rodriguez} that were included in {\it Gaia} DR1. The youth of the majority of these 19 mid-K to early-M stars was supported by their positions, relative to both the loci of main sequence (MS) stars and theoretical isochrones, in {\it Gaia} colour-magnitude and colour-colour diagrams. Surprisingly few of the GALNYSS stars included in {\it Gaia} DR1 have kinematics consistent with membership in known MGs, however \citep{2017a_Kastner}. Recent all-sky searches for nearby, young stars are beginning to identify dozens of new young stars unassociated with nearby MGs (\citealt{2019a_Schneider,2019a_Bowler}).

In the present work, we further investigate the ability of {\it Gaia} to identify nearby, young stars and to assess their MG memberships --- or lack thereof. Guided by the \citet{2017a_Kastner} study, we have used {\it Gaia} DR1 to select a sample consisting of a few hundred bright ($7 < V < 12.5$) stars with GALEX UV counterparts that are isochronally young (ages $\sim 80\,$Myr). We then used DR2 data to assess these stars' kinematics. For subsamples of these {\it Gaia}/GALEX-selected nearby young star candidates, additional archival data  (e.g., X-ray emission and Li absorption) have been compiled with which we can assess diagnostics of youth. \S\ref{sec:selection}~describes how the catalog of 146 stars was generated, and \S\ref{sec:properties}~details how additional data were obtained from available literature sources and the spectroscopic observations that were acquired for a small sub-sample of the catalog. \S\ref{sec:kinematics}~describes how Galactic kinematics were calculated as well as the kinematic tests used to test for MG membership. Results are presented in \S\ref{sec:results}, focusing on stars that are new highly probable MG members and the majority of stars that have anomalous kinematics compared with known nearby MGs. In \S\ref{sec:discussion}, we discuss the results and comment on specific areas where {\it Gaia} data have great potential to improve our understanding of young stars near the Sun. The results and conclusions are summarized in \S\ref{sec:summary}.

\section{Selecting Candidate Nearby, Young Stars from GALEX and {\it Gaia} Data}\label{sec:selection}

The present study was initiated before the release of DR2, and so relies entirely on DR1 for sample selection. However, for the analysis described in \S\ref{sec:kinematics}, we use DR2 astrometric and photometric data. This substitution is justified by the fact that, for our final sample of 146 objects, the absolute difference between DR1 and DR2 parallaxes is less than twice the combined error bar from DR1 and DR2 for $> 95\,$per cent of the sample stars, and this difference is never larger than $5.0\,$pc. Expanding on \citet{2017a_Kastner}, our initial selection of stars for the present study was based on crossmatching {\it Gaia} DR1 catalog entries with the GALEX All-sky Imaging Survey (AIS) point source catalog, but without the additional proper motion constraints used by \citet[][]{2013a_Rodriguez}\footnote{Here, as in \cite{2013a_Rodriguez}, we select stars on the basis of presence of NUV photometry in the GALEX AIS catalog, without regard to presence of FUV, because of the far more complete NUV coverage of the sky in the GALEX AIS data.}. We adopted a cross-matching radius of $3''$ to associate {\it Gaia} DR1 entries with NUV photometry from the GALEX AIS and, subsequently, near-IR ($JHK_{\rm s}$) and mid-IR ($W1-W4$) photometry from 2MASS and WISE, respectively, using the Vizier crossmatch service\footnote{\url{http://cdsxmatch.u-strasbg.fr/xmatch}}. This cross-matching exercise generated a catalog with 715,773 objects. 

\begin{figure}
\centering
\includegraphics[width=0.45\textwidth,angle=0]{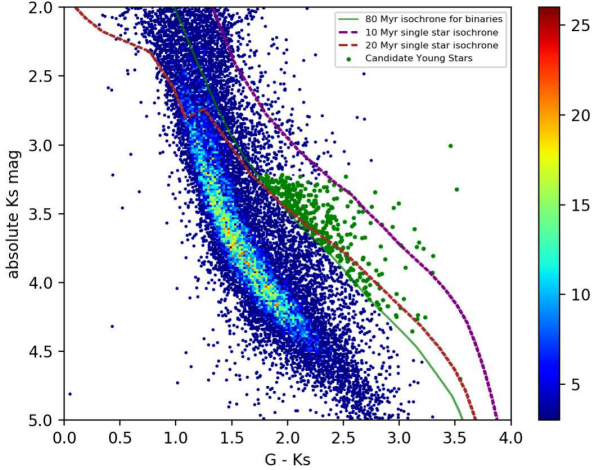}
\caption{A $K_{\rm s}$ versus $G-K_{\rm s}$ colour-magnitude diagram (density plot) of GALEX UV-selected DR1 stars, with positions of 376 young star candidates indicated as green circles. Theoretical pre-MS isochrones from T11 for ages of 10 and $20\,$Myr as well as $80\,$Myr ``binary stars'' are overlaid; i.e., the $80\,$Myr isochrone has been adjusted upwards by $0.75\,$mag, to simulate the positions of equal-components binaries of that age.}
\label{figure:K_GK_CMD}
\end{figure}

We then selected stars within $125\,$pc (i.e., parallaxes $\pi \geq 8\,$mas) that lie significantly above the MS according the models of \citet[][hereafter T11]{2011a_Tognelli}. Specifically, the T11 isochrones were used to select the subset of stars that appear younger than $80\,$Myr --- even if equal-components binaries \citep[see, e.g.,][]{2017a_Kastner} --- in a $M_{K_{\rm s}}$ versus $G-K_{\rm s}$ colour-magnitude diagram (Figure~\ref{figure:K_GK_CMD}). The evolutionary models of \citet{2015a_Baraffe} and those of T11 agree to within a few tenths of a magnitude at $80\,$Myr for K and early-M type stars, with the T11 models consistently predicting older ages for low-mass stars \citep{2017a_Kastner}. For this reason, the T11 models are a more conservative choice for selecting stars younger than $80\,$Myr.

To further limit the sample size, we then selected only stars lying below (less luminous than) and redward of the T11 $1.0\,M_{\odot}$ evolutionary track. No lower limit on mass was imposed, although the use of the Tycho catalogue to construct the DR1 catalog imposes a magnitude limit of $V \sim 12$, which should result in a sample dominated by young K and early M dwarfs (see \S\ref{sec:spectral_types}).  Figure~\ref{figure:K_GK_CMD} shows all of the Gaia DR1 and GALEX crossmatched stars, highlighting the 376 stars that we selected on the basis of lying above the 80 Myr equal-components binary isochrone and below and redward of the $1.0\,M_{\odot}$ pre-MS track.

\begin{figure}
\centering
\includegraphics[width=0.45\textwidth,angle=0]{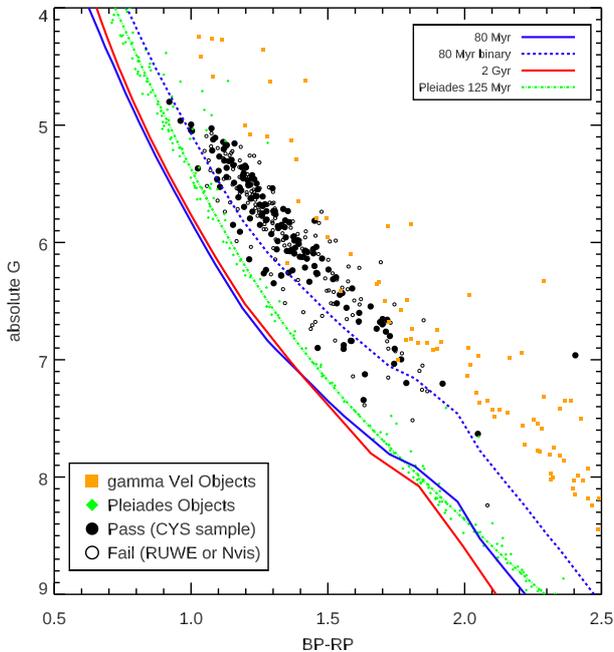}
\caption{Absolute $G$ versus $G_{\rm BP} - G_{\rm RP}$ CMD, where open circles denote the 159 objects that satisfy all membership criteria except for the DR2 quality flags, and the other 146 objects that do satisfy these criteria --- and, hence, constitute the final CYS sample --- are displayed as black filled circles. Solid green diamonds and solid orange squares represent Pleiades ($125\,$Myr) and $\gamma$~Vel ($18-21\,$Myr) members, respectively. The green dashed curve is the quintic polynomial used to describe the Pleiades single sequence. The red and blue solid lines map the 2\,Gyr and 80\,Myr PARSEC isochrones, respectively.}
\label{figure:G_BP_RP_CMD1}
\end{figure}

Since chromospherically-active stars can vary by several tenths of a magnitude in a given photometric band, and DR2 and 2MASS data are separated by observational baselines $> 10\,$years, we further vetted the candidates that lie above the $80\,$Myr equal mass binary isochrone in Figure~\ref{figure:K_GK_CMD}~via a CMD based purely on DR2 photometry. The resulting $M_{\rm G}$ versus $G_{\rm BP}-G_{\rm RP}$ CMD is shown in Figure~\ref{figure:G_BP_RP_CMD1}, overlaid with isochrones for single and equal-mass binary stars of age $80\,$Myr and a $2\,$Gyr single-star isochrone to represent the MS \citep[where these isochrones have been generated from the PARSEC 3.1 evolutionary models;][]{2012a_Bressan}.

Given the significant issues that have been raised by the astronomy community regarding the calibration of DR2 passbands with isochronal models \citep[e.g.,][]{2018a_Gagne}, we also compare our candidates with an empirical isochrone constructed for Pleiades cluster members \citep[age $= 125 \pm 9\,$Myr,][]{1998a_Stauffer,2014a_Melis}. We collected DR2 photometry for 234 Pleiades candidates with $G < 15$ that have $> 99$\,per cent membership probability in \citet{2018a_Olivares}. To represent the single sequence for the Pleiades, we fit a quintic polynomial, removed all objects that lie $> 0.25\,$mag above the fit, and again fit with the same polynomial to the surviving members; the resulting sequence is represented by the green dashed curve in Figure~\ref{figure:G_BP_RP_CMD1}. Restricting candidates to those that lie above the empirical single-star Pleiades curve reduces the sample to 336 stars. The polynomial coefficients for the fit to the Pleiades single sequence are $P0 = 2.782, P1 = -6.598, P2 = 20.528, P3 = -16.178, P4 = 5.548$ and $P5 = -0.708$, with a dispersion in the fit $\sim 0.020\,$mag. Solid orange squares in~Figure~\ref{figure:G_BP_RP_CMD1} represent a sample of members in the $\gamma$~Vel cluster ($18-21\,$Myr, \citealt{2014a_Jeffries}). The $\gamma$~Vel sample are located in similar regions of the CMD as our target selection, confirming that our selected stars may be genuinely as young as $\gamma$~Vel members. We further require that stars have a 2MASS $K_{\rm s}$ quality flag (Qflag) of value A, B or C, and have artefact contamination (Cflag), extended source contamination (Xflag), and asteroid or comet association (Aflag) flags all set to zero, leaving 305 objects in the sample.

Finally, at the recommendation of the referee we employ a final requirement that all stars have at least 8 DR2 visibility periods and for which the ``re-normalised unit weighted error'' (RUWE) is $\leq 1.4$, as suggested in \cite{2018b_Lindegren}, using the normalisation factors provided by Gaia's Data Processing and Analysis Consortium (DPAC). This cut helps filter contamination and astrometric effects from binary stars, and stars with poorly calibrated 5-parameter astrometric solutions.

The foregoing selection criteria resulted in the sample of 146 stars that are highlighted in Figure~\ref{figure:G_BP_RP_CMD1} as filled black circles and are listed in Table~\ref{table:Sample}, and are herein referred to as the candidate young star (CYS) sample. These CYSs are more or less uniformly distributed across the sky, with the exception of the GALEX Galactic plane avoidance zone. The candidate sample includes a small number of previously identified MG members (see \S\ref{sec:mg_candidates}) --- including TW Hya, the namesake of the $\sim$10 Myr-old association whose identification spawned the wider search for MGs and their members \citep{1997a_Kastner,2004a_Zuckerman,2008a_Torres}. All object designations in the text are 2MASS identifiers and should be prefixed by ``2MASS J''.

Both figures\,\ref{figure:K_GK_CMD}~and~\ref{figure:G_BP_RP_CMD1} demonstrate that our CYS sample are clearly not well described by standard MS isochrones, in terms of their CMD positions, regardless of whether one refers to the theoretical or empirical (Pleiades) curves. It is furthermore apparent that the 159 stars that fail one or both DR2 quality flags (see above) occupy the same domain of Gaia CMD space (Figure\,\ref{figure:G_BP_RP_CMD1}) as the 146 ``survivors''. We comment further on the 159 ``rejected'' stars in $\S$\ref{sec:discussion}. We note that use of DR2 instead of DR1 as the initial source of cataloged stars to be cross-matched with GALEX~AIS would result in a sample of stars approximately a factor 9 larger than that listed in Table~\ref{table:Sample}. Analysis of this far larger sample of $\sim 1500$ stars is beyond the scope of this paper.

\input{T_Sample}

\section{Properties of the Candidate Stars}\label{sec:properties}

\subsection{Spectroscopic data}\label{sec:spectra}

To assess the youth and evolutionary status of the CYSs closest to Earth, we obtained spectra for a small but presumably representative subsample of 24 stars in the brightness range $7.5 < G < 11.6\,$mag, the majority of which have distances $d < 50\,$pc. These 24 stars were randomly drawn from the sample of 305 stars that we compiled before application of DR2 data quality flags, since at the time of observing we had not enforced any DR2 quality criteria. Twelve stars are among the ``surviving'' CYS sample of 146 stars, and 12 are from the group of 159 that fail DR2 quality flag tests. This allows us to check on systematic differences between the spectra of DR2 data quality ``rejects'' versus ``survivors''.

Medium-resolution echelle optical spectra ($R \sim 18,000$ at $5000\,$\AA) were obtained for three candidate stars on 4 consecutive nights commencing June 30, 2017 using the echelle spectrograph on the 2.1-m telescope at the San Pedro M\'artir observatory (SPM), M\'exico. This instrumental setup has a 2-pixel resolution of $\sim 17\,{\rm km\,s}^{-1}$. Over 15 echelle orders, the wavelength range of $\lambda\lambda$ ($4950-7000\,$\AA) easily covers the H$\alpha$ and Li feature at 6562.8 and 6707.7\AA, respectively. An additional 12 stars were observed with the Wide-Field Spectrograph (WiFeS) on the Australian National University 2.3-m telescope at Siding Spring Observatory \citep[$R \sim 7,000$ at $7000\,$\AA,][]{2007a_Dopita} over two consecutive nights commencing September 1, 2018, covering a wavelength range almost identical to the spectra obtained at SPM. Finally, 11 high-resolution spectra were obtained with the Magellan Inamori Kyocera Echelle on the Magellan telescope at the Las Campanas Observatory with $1 \times 1$ binning, \citep[MIKE, $R \sim 65,000$ in the red arm, using the 0.35'' slit][]{2003a_Bernstein} on the night of 24 March, 2019 with an average seeing of 0.45''. The MIKE spectrograph covers 34 echelle orders between $\sim 5000-9000$\,\AA, with a sampling units of 0.05\,\AA/pixel.

We present our spectroscopically measured radial velocities (RVs), H$\alpha$ and Li equivalent widths (EWs) in Table~\ref{table:Spectroscopy_Observations}, and heliocentrically-corrected, normalised spectra centered around the H$\alpha$ and Li line are presented in Figure~\ref{figure:Spectra}, where labels ``R'' and ``S'' correspond to either a rejection or a survival from the DR2 astrometric quality cuts, respectively. As noted, the representative spectroscopic sample were observed irrespective of their DR2 flag status, and we do not find any distinction between spectral characterstics of the ``rejects'' and ``survivors''. Only the 12 stars that survive the DR2 quality criteria are used for further analysis.

\subsubsection{Radial velocities}\label{sec:rv_measurement}

To avoid regions unsuitable for calculating RVs the ANU long-slit spectra are separated into bins 50\AA~wide to simulate echelle orders. For all spectra, relative RV shifts are calculated by fitting the cross-correlation function (CCF) using either a Gaussian (for narrow CCFs) or a parabola (for broad CCFs), and locating the pixel shift corresponding to the maximum height, using the IRAF program {\sc fxcor} \citep{1979a_Tonry}. We measure the RV shift for each echelle order and iteratively remove any outliers that are $> 2\sigma$ from the mean, avoiding orders that contain telluric contamination or very gravity-sensitive features. The final RV for all observations is weighted using the Tonry-Davis `R' factor, which characterises the height of the CCF peak compared to the full-width at half maximum.

For SPM and ANU observations, heliocentric RVs were calculated by cross-correlating the science spectra with RV template stars (observed on the same night and applying heliocentric velocity corrections to both the template and target star) within half a spectral type. For the MIKE spectra, the RV templates are replaced by synthetic spectra \citep{2005a_Coelho} at solar metallicity and $\alpha$-enhanced chemical composition, degraded to the MIKE spectral resolution and signal-to-noise (SNR) over each echelle order. The effective temperature ($T_{\rm eff}$) of the synthetic template chosen in each RV cross-correlation corresponds to the closest match in $T_{\rm eff}$ for the target star based on $G-K_{\rm s}$. For SPM and ANU spectra, errors in RV are added in quadrature from three sources: 1) the error bar from the RV template, 2) the standard deviation in RV measurement across the spectral orders used in the cross-correlation and 3) the error from cross-correlation of all RV templates on a given night. For MIKE spectra only error source 2 is used.

\subsubsection{Equivalent widths}\label{sec:ew_measurement}

We followed the same procedure in \citet{2015a_Binks} to measure H$\alpha$ and Li EWs and uncertainties for the Li\,{\sc i}~line and corrected for a blended Fe line at 6707.4\AA, by subtracting $20{\rm (}B-V{\rm )} - 3\,$m\AA, from the measured EW \citep{1993a_Soderblom}. The Li EW uncertainties are calculated using the formulation given by \citet{1988a_Cayrel_de_Strobel}. Where no Li line is apparent, or where the Li EW is found to be less than $40\,$m\AA, we quote $2\sigma$ upper limits. For strong H$\alpha$ lines ($> 0.5\,$\AA) EW errors are assumed to be $\sim 0.1\,$\AA. To identify any potential giants (or sub-giants) in our spectroscopic sample, we assess the relative levels in the continuum around the Ca triplet (at 6102, 6122, 6162\,\AA), using the prescription in \cite{2012a_Prisinzano}. We find no evidence that any of the stars in our own spectroscopic sample are giant stars.

\input{T_Spectroscopy_Observations}

\begin{figure}
\centering
\includegraphics[width=0.45\textwidth,angle=0]{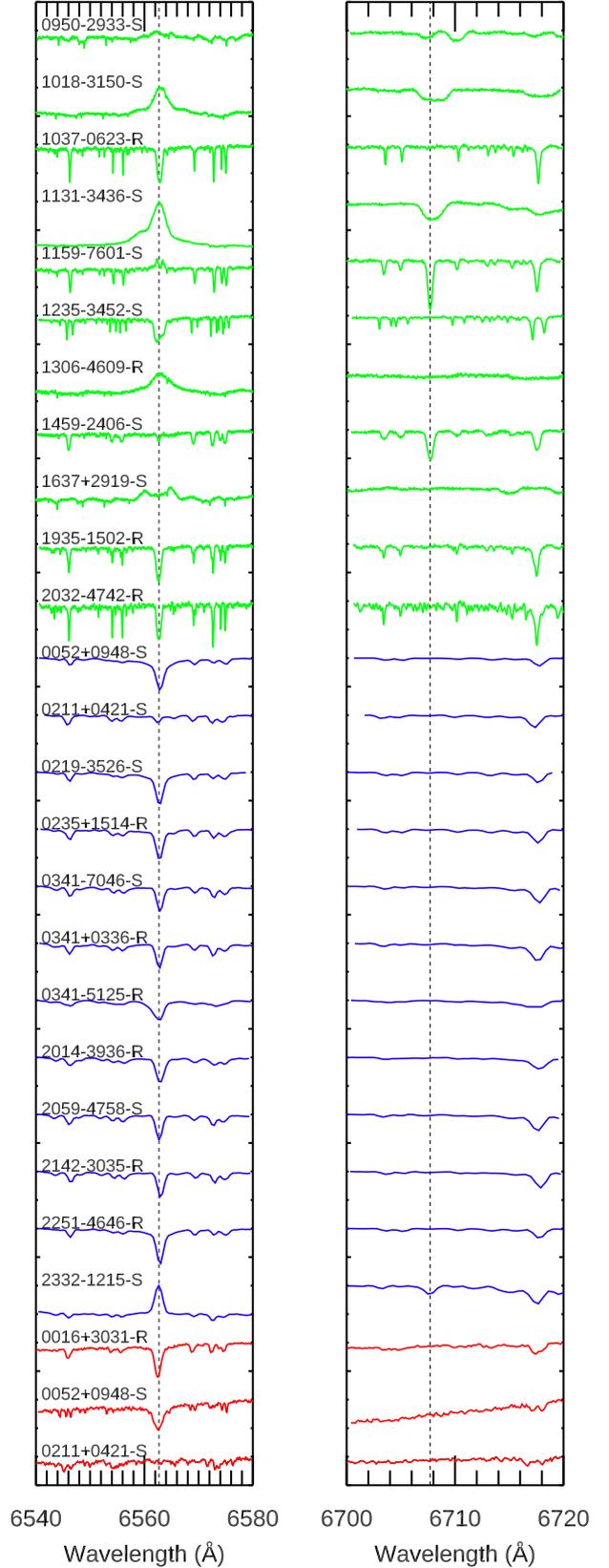}
\caption{Spectra for the candidates observed at the SPM (red), ANU (blue) and MIKE (green) around the H$\alpha$ feature at $6563\,$\AA~(left) and the Li\,{\sc i}~feature at $6708\,$\AA~(right). The name suffixes `S' and `R' indicate whether the star survives or is rejected from the CYS sample based on the DR2 astrometric criteria.}
\label{figure:Spectra}
\end{figure}

\subsection{Spectral types}\label{sec:spectral_types}

\begin{figure}
\centering
\includegraphics[width=0.45\textwidth,angle=0]{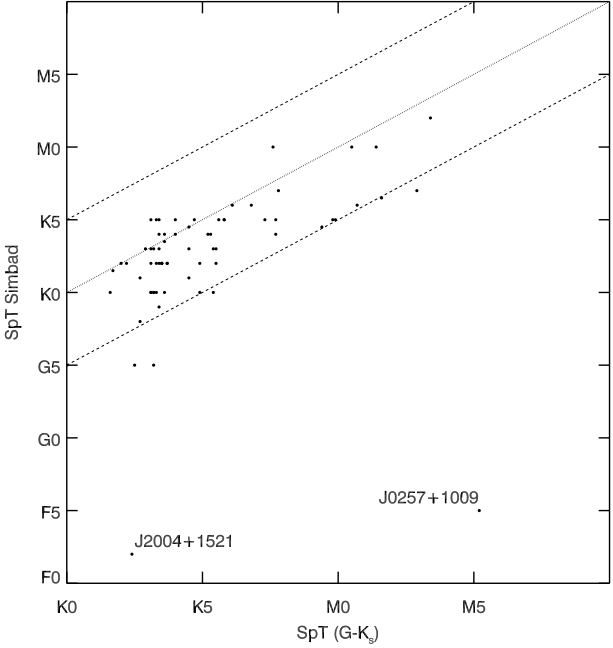}
\caption{A comparison of SIMBAD spectral types with photometric spectral types. The solid line represents unity and the two dotted lines are deviations by half a spectral class.}
\label{figure:spectral_type}
\end{figure}

We have assigned spectral types to the 146 CYSs by linearly interpolating $G-K_{\rm s}$ versus spectral type ($G$ magnitudes all from DR2) using the table provided by E. Mamajek\footnote{(herein EEM -- \url{http://www.pas.rochester.edu/~emamajek})}. We refer to these as photometric spectral types. For comparison, we find 69 stars with spectral types designated in the SIMBAD database (\citealt{2000a_Wenger}). Figure~\ref{figure:spectral_type} shows that the majority of stars with spectral types available in SIMBAD (black filled circles) agree with the photometric spectral types to within half a spectral class, although we notice that SIMBAD spectral types are systematically approximately 2 spectral sub-classes earlier than the photometric spectral types.

We find 4 examples in SIMBAD where the luminosity class is given as either a sub-giant or giant. Two stars (labelled in Figure~\ref{figure:spectral_type}) have SIMBAD spectral types $\sim 2$ entire classes earlier than the photometric spectral types. Since both stars are within 100\,pc we do not expect this discrepancy to be due to reddening effects; rather, it appears the SIMBAD source may not be correctly cross-matched with the DR2 source. As a cross check, for our spectroscopic sample we find good agreement with the final spectral type adopted and those derived from our own spectral type analysis using the {\sc sptclass} code \citep[][listed in column 5 of Table\,\ref{table:Spectroscopy_Observations}]{2017a_Hernandez}.

As expected, the vast majority of the stars have K spectral types, a result of the combination of the range of $G-K_{\rm s}$ over which they were selected and {\it Gaia} DR1 magnitude limits \citep[see][]{2017a_Kastner}. In total, the number of stars with spectral type G, K and M are 5, 134 and 7, respectively. Spectral types and photometric magnitudes are listed in Table~\ref{table:Sample}.

\subsection{Age diagnostics}\label{sec:age_diagnostics}

\subsubsection{NUV excess}\label{sec:nuv}

\begin{figure}
\centering
\includegraphics[width=0.45\textwidth,angle=0]{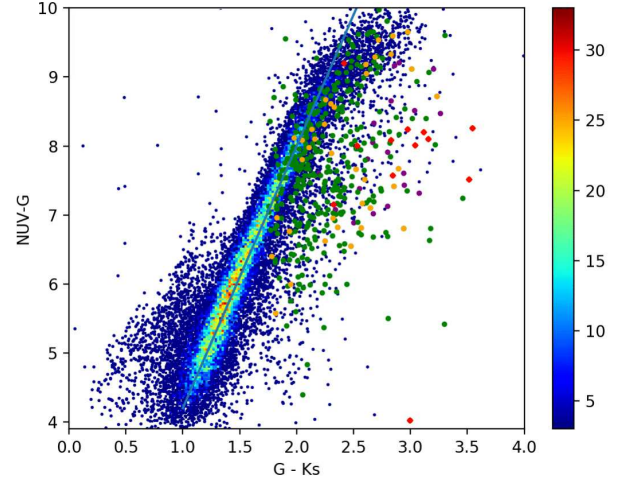}
\caption{$NUV-G$ versus $G-K_{\rm s}$ colour-colour diagram for all GALEX-selected stars in {\it Gaia} DR1, highlighting the same (376) initial DR1-selected candidates that are highlighted as green circles in Figure\,\ref{figure:K_GK_CMD} (green symbols; yellow symbols for stars within $50\,$pc) as well as the sample of (19) GALNYSS stars from \citet[][purple symbols]{2017a_Kastner} and previously identified nearby, young stars (red symbols). The green line indicates the locus of MS stars; the displacement of a star to the right of this line is indicative of the presence and strength of NUV excess.
\label{figure:NUV_Diff}}
\end{figure}

{Luminous UV is likely necessary but is clearly not sufficient to classify/identify a star as young; hence UV remains a good criterion for initial selection of young star candidates, as long it is subject to reasonable caution (and follow up).} \citet{2017a_Kastner} demonstrated that UV-selected nearby young stars generally appear below the locus of MS stars in a $NUV-G$ versus $G-K_{\rm s}$ colour-colour diagram, due to their enhanced levels of chromospheric activity and (hence) near-UV excesses. Figure~\ref{figure:NUV_Diff} confirms that the larger sample considered here adheres to this trend; the majority of the selected stars indeed lie below the MS locus. From the vertical distance between each CYS's $NUV-G$ and the main-sequence $NUV-G$ versus $G-K_{\rm s}$ line (green line in Fig.~\ref{figure:NUV_Diff}), we obtain the estimated NUV excesses, $\Delta NUV$, that are listed in column 8 of Table~\ref{table:Sample}. The fit used to provide the MS locus is $NUV-G = 3.8(G-K_{\rm s}) + 0.4$.

\begin{figure}
\centering
\includegraphics[width=0.45\textwidth,angle=0]{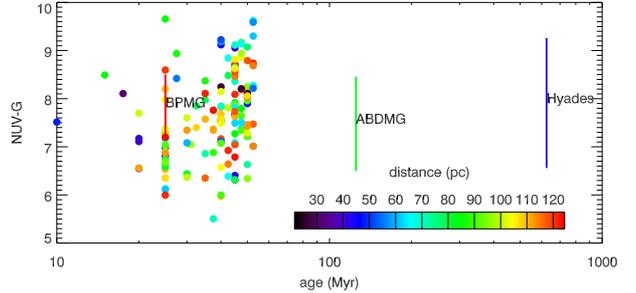}
\caption{$NUV-G$ versus isochronal age for the CYSs, with the colour coding indicating distances to individual stars. The vertical line segments indicate the means and standard deviations of $NUV-G$ for the $\beta~$Pictoris MG (BPMG), AB Doradus MG (ABDMG) and Hyades.}
\label{figure:NUV_Clusters}
\end{figure}

In Figure~\ref{figure:NUV_Clusters}, we plot $NUV-G$ versus isochronal age (and distance) for the 146 CYSs. For reference, we overplot the $NUV-G$ means and 1$\sigma$ dispersions for K stars in the $\beta$ Pic Moving Group \citep[BPMG, age $21-26\,$Myr;][]{2014a_Binks,2014b_Malo}, AB Doradus Moving Group \citep[ABDMG, age $\sim 150\,$Myr;][]{2015a_Bell} and Hyades \citep[age $650 \pm 70\,$Myr;][]{2018a_Martin}, which are measured as $7.82 \pm 0.68, 7.48 \pm 0.98$ and $7.91 \pm 1.35$, respectively. All three groups overlap within $1\sigma$ of each other in $NUV-G$. Furthermore, the corresponding mean $NUV-G$ for 217 K-type field-stars in the Gliese-Jahreiss catalog \citep{1991a_Gliese} is $8.40 \pm 1.07$, only slightly larger than (and within $1\sigma$ of) the means of the presumably younger MG samples. While there is some separation of the two populations in Figure~\ref{figure:NUV_Diff}, these statistics, along with Figure~\ref{figure:NUV_Clusters}, suggest that $NUV-G$ (or, by extension, NUV excess) is of limited utility in isolating young K-type stars from the field population, which concurs with findings in \cite{2013a_Rodriguez}.

\subsubsection{X-ray emission}\label{sec:xray}

Strong X-ray emission in K stars is another potential indicator of stellar youth; it has long been recognised that pre-MS stars typically have measured values of $\log{L_{\rm X}/L_{\rm bol}}$ in the range $-4.0$ to $-3.0$ \citep[e.g.,][Binks \& Kastner 2019, in prep; and references therein]{1997a_Kastner,2017a_Kastner}. Of the 146 CYSs, we find 57 ($39.0\,$per cent) that have ROSAT All-Sky Survey (RASS) X-ray count rates listed in either the 1RXP, 2RXS or 2RXP catalogs \citep{1999a_Voges,2000a_Voges,2016a_Boller}. 

\begin{figure}
\centering
\includegraphics[width=0.45\textwidth,angle=0]{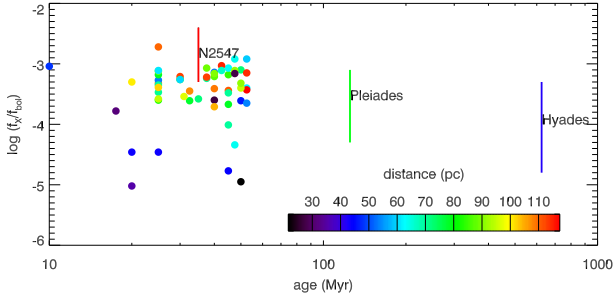}
\caption{As in Figure~\ref{figure:NUV_Clusters}, but here we display $\log{f_{\rm X}/f_{\rm bol}}$ ($=\log{L_{\rm X}/L_{\rm bol}}$) versus isochronal age and distance for the candidate stars with RASS X-ray detections. The vertical line segments indicate the means and 1$\sigma$ dispersions of three young clusters.}
\label{figure:Xray_Clusters}
\end{figure}

RASS count rates were converted to $f_{\rm X}$ as described in \citet{2016b_Kastner}, and bolometric fluxes ($f_{\rm bol}$) were estimated from the stars' spectral types and $J$ magnitudes using bolometric corrections listed in \citet{2013a_Pecaut}. The resulting plot of $\log{f_{\rm X}/f_{\rm bol}}$ ($=\log{L_{\rm X}/L_{\rm bol}}$) versus age (and distance) for the candidate stars with RASS X-ray detections is presented in Figure~\ref{figure:Xray_Clusters}, overlaid with the means and 1$\sigma$ dispersions in $\log{f_{\rm X}/f_{\rm bol}}$ for K stars in three nearby young star clusters, NGC 2547 \citep[age $= 35\,$Myr;][]{2005a_Jeffries}, Pleiades and Hyades, to illustrate the temporal change in $\log{f_{\rm X}/f_{\rm bol}}$ over the age range 35\,Myr to 650\,Myr \citep[see also][]{1997a_Kastner}. Comparison of Figures~\ref{figure:NUV_Clusters} and~\ref{figure:Xray_Clusters} provides tentative evidence that NUV emission remains elevated for K stars, even after X-ray emission begins to decline, and that both the X-ray and UV distributions may broaden after K stars arrive on the MS. These results are consistent with those of \citet{2013a_Stelzer}, who found a similar relationship for UV and X-ray emission for M stars in MGs and in the field. 
The rather low overall RASS detection rate of the CYS --- combined with the fact that the detections are dominated by the nearest stars, many of which have $\log{L_{\rm X}/L_{\rm bol}} < -4.0$~(Figure~\ref{figure:Xray_Clusters}) -- suggests that our CYS sample (Table~\ref{table:Sample}) includes a significant number -- perhaps a majority -- of stars older than Pleiades age. We consider this possibility in detail in \S\ref{sec:discussion_low_yield}.

\subsubsection{H$\alpha$ emission}\label{sec:ha}

The strength of H$\alpha$ emission provides another probe of stellar activity \citep{1999a_Hawley, 2003a_White}. It is well established that young, rapidly-rotating and (hence) active stars exhibit strong H$\alpha$ emission, which therefore also serves as an age indicator \citep{2004a_Zuckerman}. Measurements of H$\alpha$ in several open clusters of known age show that low-mass stars remain chromospherically active and retain H$\alpha$ in emission for longer than solar-type stars \citep{1995a_Reid}, such that the utility of H$\alpha$ as an age diagnostic decreases with decreasing stellar mass.

Combining the results of our spectroscopic sample with a literature search using the VizieR database, 41 stars among our final list of 146 CYS (28.1 per cent of the sample) are found to have at least one H$\alpha$ EW measurement, of which 22 are in emission. For 17 stars there are 2 or more separate H$\alpha$ EW measurements, from which we find broad agreement in every case and no stars with contradictory measurements between absorption and emission.

Our complementary spectroscopic survey (see \S\ref{sec:spectra} and Table~\ref{table:Spectroscopy_Observations}) measured H$\alpha$ EW for the first time in 6 stars. We list all H$\alpha$ EW measurements and their reference sources in columns 3 and 4 of Table~\ref{table:Spectroscopy_Data}, respectively, where for stars with 2 or more H$\alpha$ EW measurements the unweighted average is given. Individual H$\alpha$ EW measurements, and their references are available in a supplementary online table.

\subsubsection{Infrared excesses}\label{sec:ir}

Evidence of either a transitional or debris disk, in the form of a mid-IR excess due to warm dust, is a potential youth indicator, as (with notable exceptions) such objects are mostly confined the age range $\sim 10-100\,$Myr \citep[see, for example][]{2008a_Wyatt,2014a_Matthews,2018a_Hughes}. Various studies using photometry from the Wide-field Infrared Survey Explorer (WISE) have established that the general K/M field star population have WISE $W1-W4$ colours centered around $W1-W4 = 0$ with a dispersion of $\sim 0.3$, such that stars with $W1-W4 > 1.0$ are candidate debris disk hosts \citep[e.g.,][]{2012a_Schneider,2017a_Binks}. There are two objects in our sample that have $W1-W4 > 1.0$ (and are labelled in Table~\ref{table:Sample}): V1317 Tau ($04234759+2940381$), previously identified as a weak-lined T Tauri star associated with the Taurus cloud \citep{1996a_Wichmann}, and the intensively studied TW Hya ($11015191-3442170$; see Table~\ref{table:Sample}), with $W1-W4 = 1.284$ and $W1-W4 = 5.585$, respectively. Four Table~\ref{table:Sample}~stars have moderate evidence of mid/far IR excess from Spitzer-based surveys: $1235-3452$ \citep{2009a_Lawler}, $04392545+3332446$, $11594226-7601260$ \citep{2010a_Wahhaj} and TW Hya \citep[$11315526-3436272$][]{2007a_Rhee}.

\subsubsection{Lithium absorption}\label{sec:lithium}

Although theoretical models for Li depletion are strongly sensitive to surface conditions and stellar opacities, one can use empirical Li EW data in open clusters as benchmarks to distinguish stellar ages. There is a large dispersion in the Li EW distribution amongst K stars in young open clusters $< 500\,$Myr, which makes Li EW a useful age discriminant for stars of this spectral type. For stars with strong Li EW signatures ($> 100\,$m\AA), ages can potentially be resolved on the order of tens of Myr. We hence use these Li EW measurements to further constrain ages of the CYS sample.

Thirty-nine of the CYSs have at least one Li $\lambda$6708\,\AA~absorption line EW measurement, obtained either from literature sources or from our small spectroscopic survey (see \S\ref{sec:spectra}). Our spectroscopic sample provides Li EW measurements for the first time in 6 stars, of which only $1459-2406$ has a strong Li line ($= 314 \pm 4$\,m\AA). There are 11 stars that have two or more Li EW measurements, and for each of these stars the standard deviation in Li EW is always less than $\sim 30$\,m\AA\footnote{With the exception of $0950-2933$ that has a standard deviation $= 50$\,m\AA.}. For stars with 2 or more measurements, we calculate their unweighted average, which is used for all subsequent analyses. These values, and the reference sources used to obtain this value are provided in column 5 and 6 of Table~\ref{table:Spectroscopy_Data}, respectively. All individual Li EW measurements are provided in an online supplementary table.

\begin{figure}
\centering
\includegraphics[width=0.45\textwidth,angle=0]{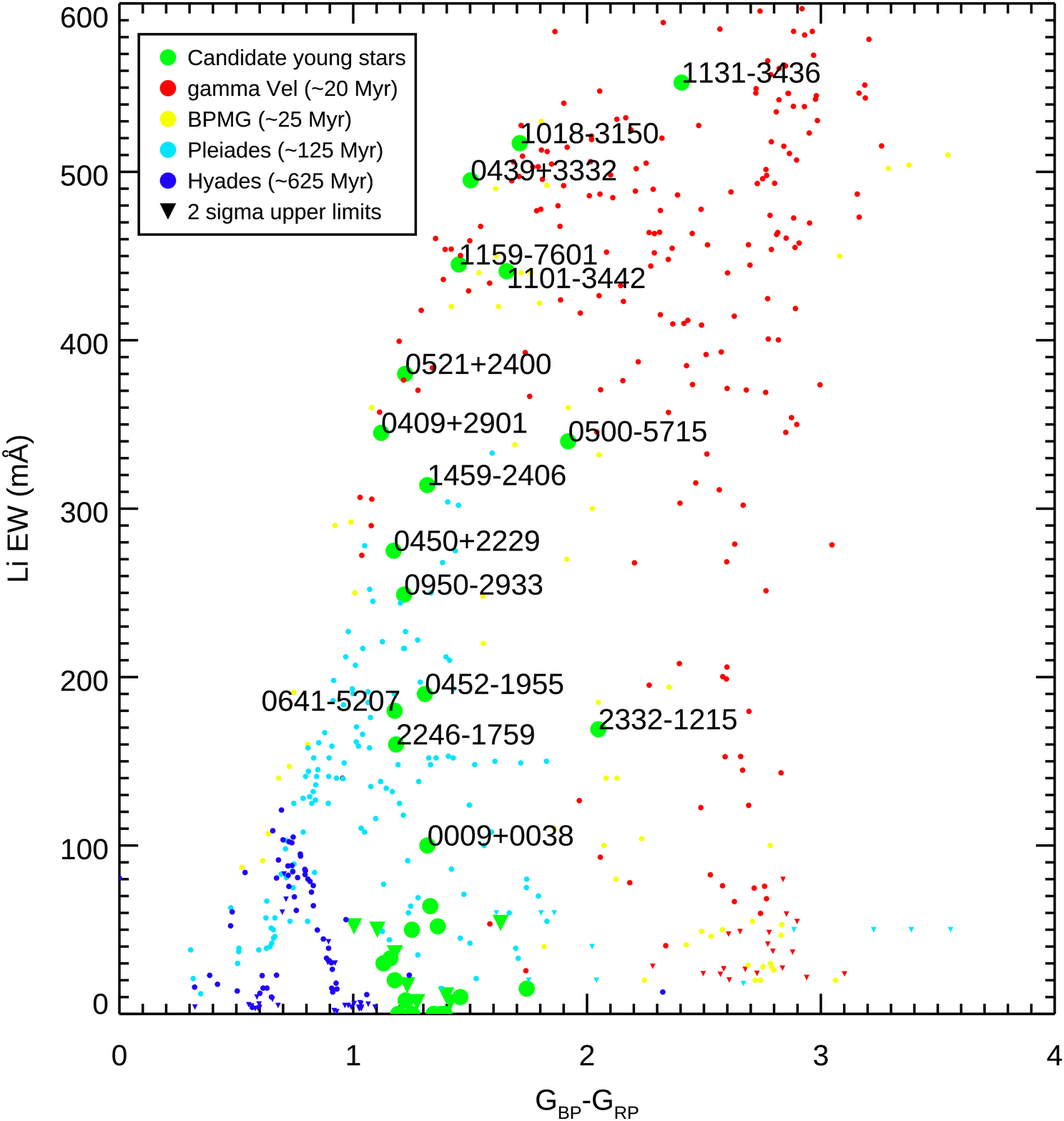}
\caption{Plot of Li EW versus $G_{\rm BP}-G_{\rm RP}$ colour for those CYSs with available Li EW measurements (Table~\ref{table:Spectroscopy_Data}) and for various young clusters. Downwards facing triangles indicate 2$\sigma$ upper-limit measurements. The labelled targets have Li age upper-limits $\leq 125\,$Myr.}
\label{figure:LiEW_BP_RP}
\end{figure}

In Figure~\ref{figure:LiEW_BP_RP} we plot Li EW against $G_{\rm BP}-G_{\rm RP}$ colour for the 39 CYSs with Li measurements overlaid on the same data for members of $\gamma$~Vel \citep[age $\sim 20\,$Myr,][]{2016a_Prisinzano}, BPMG \citep[age $\sim 25\,$Myr,][]{2016a_Messina}, Pleiades \citep{2018a_Bouvier} and the Hyades \citep[age $\sim 625\,$Myr,][]{2017a_Cummings}. This comparison suggests that 16 CYSs have Li EW consistent with ages similar to the Pleiades, or younger, and nine of these are potentially younger than 50\,Myr. Since the literature-sourced Li EW measurements are strongly biased towards stars likely to be members of young MGs, it is not surprising that the majority of Li detections in our sample of spectroscopically observed stars are previously established members of young MGs (see \S\ref{sec:Li_young_stars}). The Li-absorption-based ages inferred from Figure~\ref{figure:LiEW_BP_RP} are presented in column 6 of Table~\ref{table:Spectroscopy_Data}. Stars that have Li EWs consistent with Pleiades or younger ages are examined in more detail in \S\ref{sec:Li_young_stars}.

\input{T_Spectroscopy_Data}

\section{Kinematics}\label{sec:kinematics}

\subsection{Selecting radial velocities}\label{sec:rv_selection}

The methodology of measuring RVs for our small spectroscopic sample is described in \S\ref{sec:rv_measurement}. To supplement these measurements we performed a VizieR search for RVs for every member in our CYS sample. In total there are 102 stars (69.9 per cent) with at least one RV measurement (from our spectroscopic sample and/or the literature search), of which 27 have two or more measurements. Our spectroscopic sample measured an RV for the first time for four stars.

In some cases where an RV is reported in the literature, no error estimate is available. For these measurements we adopt a conservative error of $3.0\,{\rm km\,s^{-1}}$, which is roughly twice the median uncertainty across all measurements with a published RV uncertainty. When applying our literature search, we identified a number of stars with two or more measurements that consist of large errors combined with small errors, which would potentially smear out hard-gained, high-resolution RV measurements. Therefore we apply the following criteria: if the standard deviation amongst the error bars is larger than the standard deviation in the mean then we elect the RV measurement with the lowest error bar. If the opposite is true, then we quote our final RV as the error-weighted mean, where the final error bar is the mean amongst the component errors. If the star is flagged as a probable binary based on large variability amongst individual RV measurements (see \S\ref{sec:binarity}) then the final error bar provided is the quadrature sum of the standard deviation in the mean plus the final averaged error bar. All final RV measurements (and errors) and the source references used in each calculation are provided in columns 4 and 5 of Table~\ref{table:Kinematics}, respectively. The individual RV measurements are provided in an online supplementary table.

\subsection{Binarity}\label{sec:binarity}

The unprecedented astrometric precision in DR2 allows us to identify probable common proper motion companions (CPMCs) that previously evaded detection in past surveys (e.g., the Washington Double Star Catalog). We apply a search radius of $46.5''$. This is sufficiently wide to identify potential CPMCs out to projected separations from 2000\,AU to 6000\,AU for our nearest and most distant CYS candidates, yet minimizes background star contaminants. This search yields 43 objects ($\sim 30$ per cent of the CYS sample) with probable companions, i.e., stars with parallaxes within 3 per cent, and proper motions within 5 per cent, of a CYS.

We regard stars with CPMCs $< 3$'' as close pairs. Their proper motions may be indicative of orbital motion, however they may be prone to overlapping DR2 point spread functions (PSFs) that could distort the 5-parameter astrometric solution \citep[e.g.,][]{2018a_Kastner}. Slightly more than half of the stars with a CPMC (24 out of 43) are in close pairs, and the astometric parameters may be prone to error. However, CMPCs separated by $> 3$'' should not be problematic for DR2. Since 2MASS PSFs are significantly lower resolution than the DR2 PSFs a number of stars will have resolved DR2 photometry, but unresolved in 2MASS, thus making them appear overluminous in a DR2/2MASS CMD ($G$ versus $G-K_{\rm s}$, Figure\,\ref{figure:K_GK_CMD}). However, these stars would not appear overluminous in a ``pure'' DR2-based CMD (see discussion in $\S$\ref{sec:discussion}).

Stars in multiple systems can vary significantly in RV, which complicates attempts to determine MG status from kinematics (see \S\ref{sec:kinematics}). We attempted to flag potential binaries by identifying those stars with two or more RV measurements that vary by an amount significant enough to suggest the variations are due to orbital motion. A scoring system was applied, based on the following metrics: a score of 0 represents a star with no RV measurements, whilst a score of 1 indicates a star with just one RV value. A score of 3 is given to stars with $> 1$ measurement in which the absolute difference in the mean values of RV for all pairs is $< 5\,{\rm km\,s}^{-1}$, and the quadrature sum of the error bars for every pair is $< 5\,{\rm km\,s}^{-1}$, indicating that they are likely single stars. Stars that score 5 have $> 1$ measurement, of which at least one pair is separated by $> 5\,{\rm km\,s}^{-1}$ and their errors (added in quadrature) is $< 5\,{\rm km\,s}^{-1}$, such that the RV variability is indeed indicative of binary orbital motion. Finally stars with score 9 have $> 1$ RV measurement, however, their error bars are too large to predict their binary status. We identify 4 objects in our CYS that are likely to be binary stars based on their RV variability; these 4 constitute $\sim$\,15 per cent of the sample with two or more RV measurements.

The object types that are provided in the SIMBAD database were also used to indicate whether any objects are flagged as members of multiple systems. In total there are 55 stars (38\% of the CYS) that display evidence of binarity, either based on DR2 companions, our binary scoring, or SIMBAD indication (37.7 per cent of the CYSs). The multiplicity fraction for our predominantly K-star sample fits entirely in the progression observed in volume-limited stellar multiplicity samples of both G-dwarfs \citep[$44 \pm 3$ per cent][]{2010a_Raghavan} and early M-dwarfs \citep[$31.1 \pm 3.4$ per cent, section 6.1 in][]{2019a_Winters}. It is entirely possible that our search may miss potential MG candidates among these 55 stars, or provide false positives, because of RV variation due to binarity. The binary scores used in this work and any evidence pertaining to binary stars from SIMBAD are presented in column 5 of Table~\ref{table:Kinematics}. We briefly discuss the binary status for stars of notable interest in \S\ref{sec:results}.

\input{T_CPMB}

\subsection{Determining $UVW$ space velocities}\label{sec:uvw}

Galactic space velocities ($UVW$) and their errors are calculated using positions, proper motions, RVs, and parallaxes (and their associated errors) following the prescription in \citet{1987a_Johnson}. All 146 {\it Gaia} DR1-selected objects in our CYS sample have position, proper motion, and parallax data available in the DR2 catalog. The median errors are $0.0411\,{\rm mas}$, $0.0847\,{\rm mas\,yr}^{-1}$ and $0.0400\,{\rm mas}$, respectively, where the positional and proper motion errors are the quadrature sum of the error components in right ascension and declination. Since the parallax signal-to-noise ratios are never $< 10$ we simply treat distances as the parallax inverse. 

\begin{figure*}
\centering
\includegraphics[width=0.8\textwidth,angle=0]{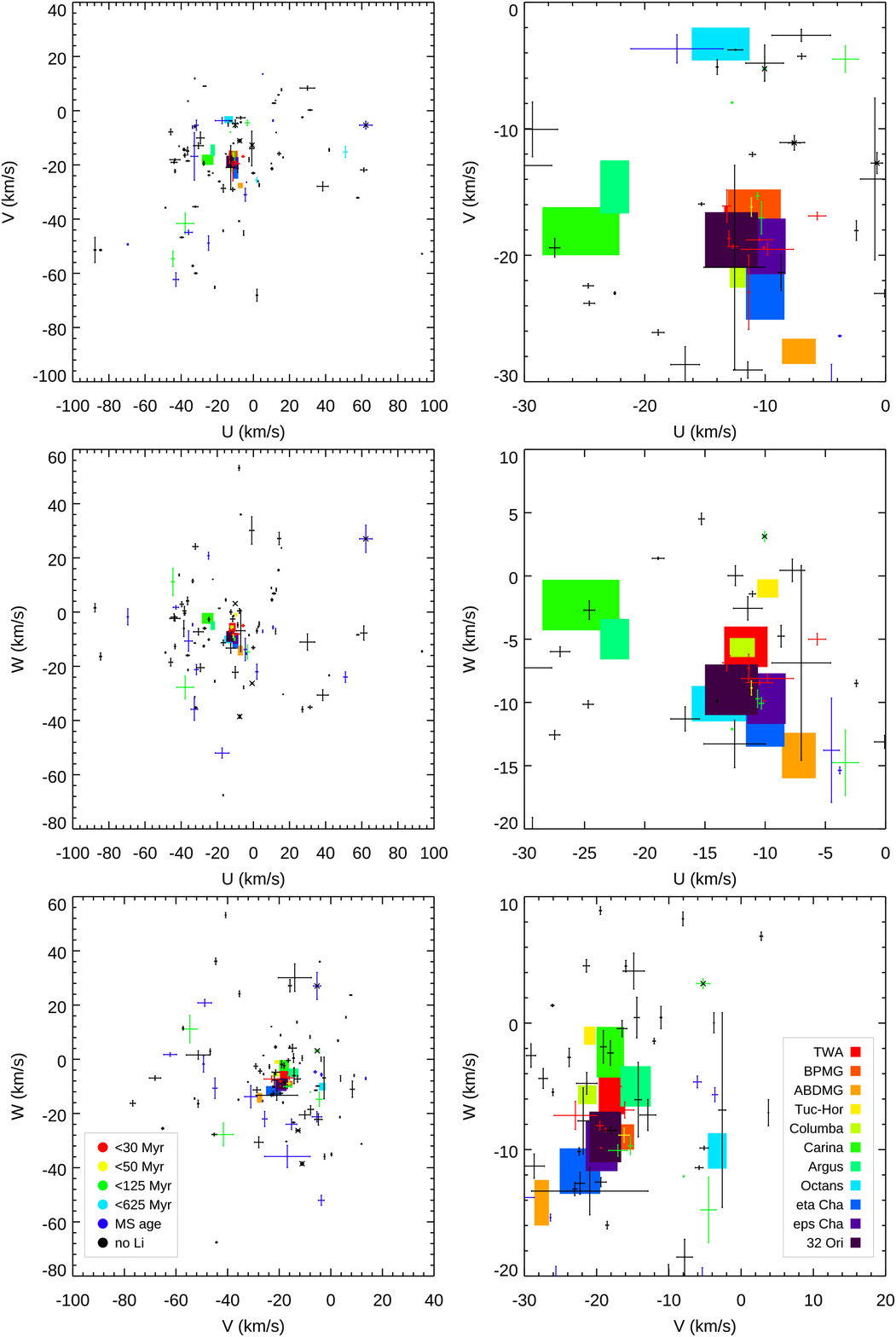}
\caption{$UVW$ space velocities of the 102 nearby young star candidates with an RV measurement, along with the 1$\sigma$ ranges of 11 well-established MGs \citep{2018a_Binks} indicated. The left panels encompass the $UVW$ range of the entire sample, whereas the right panels provide zoomed-in views of the MG centroids.}
\label{figure:UVW}
\end{figure*}

For the 44 stars that have no RV measurement, $UVW$ are calculated over a range of $-100 < {\rm RV (km\,s^{-1}}) < + 100$ in steps of $0.5\,{\rm km\,s^{-1}}$. The $UVW$ for each star are listed in columns 7, 8 and 9 of Table~\ref{table:Kinematics}, where the $UVW$ for stars without an RV are displayed by their extrema from the RV range. Figure~\ref{figure:UVW} depicts the $UVW$ distribution for the stars with RV measurements, with the 1$\sigma$ extents of 11 well known nearby MGs overplotted \citep[$UVW$ means and 1$\sigma$ dispersions are taken from the covariance diagonals quoted in table 7 of][]{2018a_Gagne}.

\input{T_Kinematics}

\subsection{Kinematic tests for MG membership}\label{sec:kinematic_tests}

Measuring Galactic kinematics and comparing to those of known young MGs is the one of the primary methods to confirm the youth of CYSs and to test their memberships in MGs. If a CYS appears to be co-moving with a known MG, it is a potential member and could therefore have an age co-eval with the proposed host group. To establish whether there exists a subset of our 146 candidate stars that are possible or established MG members, we applied the following four kinematic and distance-based membership tests.

\begin{enumerate}

\item A $\chi^{2}_{\rm MG}$ test, as described in \citet{2011a_Shkolnik} and \citet{2015a_Binks}, where $\chi^{2}_{\rm MG}$ is defined as:

\begin{center}
$\frac{(U_{*}-U_{\rm MG})^2}{\sigma_{U_{*}}^2 + \sigma_{U_{\rm MG}}^2} + \frac{(V_{*}-V_{\rm MG})^2}{\sigma_{V_{*}}^2 + \sigma_{V_{\rm MG}}^2} + \frac{(W_{*}-W_{\rm MG})^2}{\sigma_{W_{*}}^2 + \sigma_{W_{\rm MG}}^2}$
\end{center}
and is required to be $\leq 3.78$. Such a threshold rejects the null hypothesis with 95 per cent confidence. The $UVW$ data for MGs used in this analysis are from table 7 of \citet{2018a_Gagne}.

\item A kinematic distance test comparing the measured distance and the expected distance if the object were a member of the cluster, in which the difference between the measured distance and the expected distance must be less than 10\,pc.

\item The star must have a RV measurement within $5\,{\rm km\,s}^{-1}$ of the expected RV were it a member of a given MG.

\item The star must have a mean distance measurement within the 3$\sigma$ dispersion of the proposed MG, where the mean distance and 1$\sigma$ dispersions are the quadrature sums of $XYZ$ and their dispersions from table 7 of \citet{2018a_Gagne}.

\end{enumerate}

These criteria are applicable for {\it all} objects with a RV measurement in the CYS list, and for stars without RV measurement all criteria except for the RV test can be applied.

We tested a reasonable range of RVs for the stars in the CYS sample that have no measured RV, and found that one of them, $2223+3231$, could be within  $\chi^{2}_{\rm MG} \leq 3.78$ of a MG if the actual RV lies within a small range given in Table~\ref{table:MG_Candidates}. In addition to the kinematic criteria, one must ensure that the measured age of the star is at least consistent with the age range of the proposed host MG. Individual stars will be discussed in the following section.

\section{Results}\label{sec:results}

\subsection{Candidate MG members}\label{sec:mg_candidates}

After applying the criteria described in \S\ref{sec:kinematic_tests}, we find a subset ($\sim 15\,$per cent) of CYSs that are potential MG members, based on either our kinematic criteria, previous membership assignments in the literature, or both. In this section we assess the evidence for MG membership based on data collected in this work, or from previous literature, and provide our diagnosis for each object.

In Table~\ref{table:MG_Candidates} we list the 21 potential MGs members identified among the 146 CYSs. The table is split into three sections. For each candidate MG member we utilise the BANYAN$\Sigma$\footnote{available at \url{http://www.exoplanetes.umontreal.ca/banyan}} probability code, which calculates the probabilities of membership in 29 young associations within 150\,pc of the Sun based exclusively on astrometric and kinematic data.

All MGs for which BANYAN$\Sigma$ membership probabilities are $> 10$ per cent are listed in column 10. We adopt such a low BANYAN$\Sigma$ probability threshold so as to attempt to identify new candidate MG members that could lie in tails of the probability $UVW$ distributions, and also because we do not strictly employ the $XYZ$ positions as a membership criterion. The top two sections list stars that pass all appropriate kinematic membership criteria in \S\ref{sec:kinematic_tests} with reference to any of the eleven MGs considered in this work. There are 13 stars with an RV measurement in the top section of Table~\ref{table:MG_Candidates}, and one star without a measured RV in the middle section of the Table. The membership status of this star is more ambiguous because it requires an RV measurement within the range quoted in column 5 to pass the RV criterion from \S\ref{sec:kinematic_tests}. 

The bottom section of Table~\ref{table:MG_Candidates} lists data for 7 stars that have evidence for MG membership, either from literature sources or a BANYAN$\Sigma$ membership probability $> 10$ per cent, but fail the $\chi^{2}_{\rm MG}$ criterion. The fact that previous MG searches identified some of our candidates as members that our selection criteria omitted, and vice versa, presumably reflects the way membership criteria are defined, the $UVW$ velocity distributions used for given MGs, and/or the different data sets used in the selection process. We finally note that, although none of the following objects in this section were revealed to be binaries based on RV variation, binarity may nonetheless affect the kinematics of {\it any} star in this work.

\input{T_MG_Candidates}

In the following we briefly discuss relevant factors regarding MG membership status for each star in Table~\ref{table:MG_Candidates} and provide our assessment of the most likely MG membership (if any), beginning with objects in the top section of the Table (the abbreviated forms for each MG are listed in the table notes at the bottom of Table~\ref{table:MG_Candidates}). The final designated membership status for each object is provided in the final column of Table~\ref{table:MG_Candidates}.

\subsubsection{Stars with a measured RV that pass all kinematic criteria}

The following 7 stars satisfy all kinematic criteria for membership to a given MG, and there is either literature pertaining to MG membership and/or a BANYAN$\Sigma$ probability $> 10$ per cent for the same proposed group.

{\bf 05004714--5715255:} There is a kinematic match to both BPM and TWA, although the match with BPM is much better in each membership criterion, and its sky position is inconsistent with TWA membership. There are both many listings for this star as a BPM member in the literature and a near-unity probability predicted from BANYAN$\Sigma$. The Li EW measurement is in agreement with the pattern observed amongst similar spectral type BPM members, thus we confirm its status as a {\bf member of BPM}.

{\bf 10182870--3150029:} There are two kinematic matches, which are COL and TWA. Whilst previous literature suggests this target is a TWA member, the kinematic match in our analysis is better for COL. The Li EW measurement agrees with both MGs, albeit only marginally with COL. We therefore cannot discriminate between the two potential host MGs (if indeed there is one) and designate the star as a {\bf possible member of TWA or COL}.

{\bf 11015191--3442170:} This object is {\bf TW Hya}, the eponymous TWA member \citep{1997a_Kastner}. 

{\bf 11315526--3436272:} Identified as {\bf TWA 5}, this object is another of the original five members of the TWA \citep{1997a_Kastner} and, hence, also serves as confirmation of our young-star candidate selection and MG membership methodologies. Specifically, we have 2 kinematic matches (to TWA and BPM) but TWA provides the best $\chi^{2}$ match and BANYAN$\Sigma$ predicts a 99.9 per cent membership probability to TWA. The strong Li EW line suggests the star's age $< 20\,$Myr, consistent with stars of similar spectral type that have subsequently been identified as TWA members.

{\bf 11594226--7601260:} Whilst there are 2 kinematic matches, to EPS and THO, the best $\chi^{2}$ match, by far is to EPS, and has previously been categorised numerous times as an EPS member since the original designation by \cite{2000a_Mamajek}. The BANYAN$\Sigma$ code predicts a 99.9 per cent membership probability to EPS, and the measured Li EW is consistent with this membership. We therefore (re)assign this star as most likely a {\bf member of EPS}.

{\bf 12151838--0237283:} There is one single kinematic match with ABD, and several literature sources suggest it as a candidate ABD member. There is a corresponding 19.7 per cent BANYAN$\Sigma$ membership probability. Whilst there is no Li EW measurement available for this star, it has relatively strong X-ray activity ($\log f_{\rm X}/f_{\rm bol} = -3.65$), notionally consistent with stars of ABD age. We suggest that this star requires further analysis, particularly a Li EW measurement. We concur with previous work and propose this star as a {\bf candidate member of ABD}.

{\bf 23323085--1215513:} This object has been identified several times as a bonafide member of BPM. We find only one kinematic match, which is also with BPM, and the BANYAN$\Sigma$ code provides a corresponding membership probability of 99.9 per cent. The final Li EW measurement, which includes one measurement obtained in this work is consistent with membership, therefore we confirm its status as a {\bf member of BPM}.

The following 2 stars, also listed in the top section of Table~\ref{table:MG_Candidates}, have membership assignments in literature sources that do not match with ours.

{\bf 04090973+2901306:} We find a potential match with THO or COL. The sky position of this star place it outside the region of sky occupied by most known THO members. It is also likely too distant to be a COL member. There are several literature sources that claim the object is a member of TAU. Despite the low BANYAN$\Sigma$ membership probability for TAU of 13.7 per cent, the distance, sky position and Li EW are all consistent with constituents in TAU. We conclude that this object is indeed most likely a {\bf member of TAU}.

{\bf 05214684+2400444:} This objects passes all kinematic criteria for both THO and EPS, whilst BANYAN$\Sigma$ predicts an almost unity probability of membership to 118TAU. The large Li EW suggests the star is $< 30\,$Myr, consistent with all three groups. The sky position definitively rules out membership with EPS, and membership to THO is marginally ruled out as the target is $\sim 5^{\circ}$ further north than the most northernly THO members in \citet{2017a_Bell}, whereas it is very close to the centroid of 118TAU. We therefore assign this object as a {\bf member of 118TAU}, which agrees with the recent designation by \cite{2019a_Bowler}.

FInally, the following four objects listed in the top section of Table~\ref{table:MG_Candidates} have no previous evidence of MG membership in the literature or any BANYAN$\Sigma$ probabilities $> 10$ per cent.

{\bf 04392545+3332446:} There are kinematic matches to EPS, THO and TWA, and a strong Li EW measurement and strong X-ray activity, and short rotation period (= 2.418 days, \citet{2006a_Watson}) are consistent with the young ages for each of these MGs. However, the sky position conclusively rules out membership in any of these groups. Several publications posit that the star is a member of TAU, despite a BANYAN$\Sigma$ of only 0.1 per cent. The largest difference in velocity is in the $V$ coordinate, both in standard deviations ($2.04\sigma$) and numerically ($\left|\Delta V\right| = 10.2\,{\rm km\,s}^{-1}$). Given the lack of alternative MGs, we tentatively designate this object as a {\bf member of TAU}.

{\bf 09503676--2933278:} We find a kinematic match with THO. The measured Li EW is consistent with membership, while the sky position is inconsistent. There are no literature sources found pertaining to a membership status, despite its inclusion in the catalog input to the SACY search \citep{2006a_Torres}. For present purposes we classify this star as a {\bf young field star}.

{\bf 11212188--4736028:} Whilst we find a kinematic match with THO, the sky position is inconsistent with the group. The object is a K2\,{\sc III} + K2\,{\sc III} eclipsing binary, which may describe the star's overluminosity compared to the CMD MS (as opposed to youth). We assign the membership status as an {\bf indeterminate field star}.

{\bf 21351099+3402313:} Only one kinematic match is found, with OCT. There are no literature references to this target, and no specific additional youth evidence. We tentatively suggest this object as a {\bf potential member of OCT}.

\subsubsection{One star without a measured RV}

{\bf 22233510+3231182:} There is a potential match with OCT for this object, and coincides with the required distance for membership. No literature references are found for this source. One measurement of H$\alpha$ is found in absorption and no additional youth indicators were identified. Thus we suggest this star is a {\bf potential member of OCT}, pending spectroscopic confirmation.

\subsubsection{Stars classed as MG members in literature, missed in our search}

Finally, there are 7 stars listed in Table~\ref{table:MG_Candidates} (bottom section) that have previously been assigned MG memberships in the literature and/or have BANYAN$\Sigma$ membership probabilities $> 10$ per cent, but that failed one or more of our kinematic tests (\S~\ref{sec:kinematic_tests}).

{\bf 04053103--0216257:} Although the DR2 parallax distance for this star is consistent with the range for the Hyades, this object is somewhat offset in from the cluster in sky position. It has a somewhat indeterminate BANYAN$\Sigma$ HYA membership probability of 48.9 per cent. Therefore, we tentatively assign it the status of a {\bf possible Hyades member}.

{\bf 04181077+2317048:} The BANYAN$\Sigma$ membership probability for HYA is high (90.2 per cent), and there are several publications that recognise it as a member. Therefore, although it does not meet our kinematic criteria, it Is a probable {\bf member of HYA}.

{\bf 04500019+2229575:} The BANYAN$\Sigma$ probability is 99.7 per cent for TAU and its distance is consistent with TAU membership. Several publications claim its membership to TAU. The age based on Li EW is marginally consistent, so --- although it does not meet our kinematic criteria --- we concur that this star is a {\bf member of TAU}.

{\bf 04524951--1955016:} This object has beeen assigned THA membership status in the literature, but broadly fails our kinematic criteria. Furthermore, it has a Li EW consistent with an age similar to that of the Pleiades. We suggest it is unlikely to be a THA member, and is instead a {\bf young field star}.

{\bf 14590325--2406318:} BANYAN$\Sigma$ predicts a 94.6 per cent probability of membership with UCL. This object narrowly passes the $\chi^2$ test and distance criterion for BPM, however the distance is far beyond the distance domain of BPM. Our MIKE spectrum for this object (see Figure~\ref{figure:Spectra}) reveals, for the first time, a strong Li EW line and thus we confirm this object as a {\bf highly likely member of UCL}.

{\bf 22132028+8445372:} There is a 66.3 per cent probability of membership to ARG predicted by the BANYAN$\Sigma$ code. This object does not have any recorded RV and we could not find a corresponding RV value that would provide a $\chi$-squared match to ARG. There are no spectroscopic youth indicators for this object, therefore we designated this for the present as a {\bf field star} awaiting spectroscopic youth confirmation.

{\bf 23005681+3713528:} This object has a high ABD membership probability (97.5 per cent) from BANYAN$\Sigma$, but narrowly fails the $\chi^{2}$ test (= 5.16). It has previously been identified as a multiple system in the ABD MG, and we concur that it is most likely a member of ABD.

\subsection{Stars within 50\,pc not associated with MGs}\label{sec:50pc}

\begin{figure*}
\centering
\includegraphics[width=0.8\textwidth,angle=0]{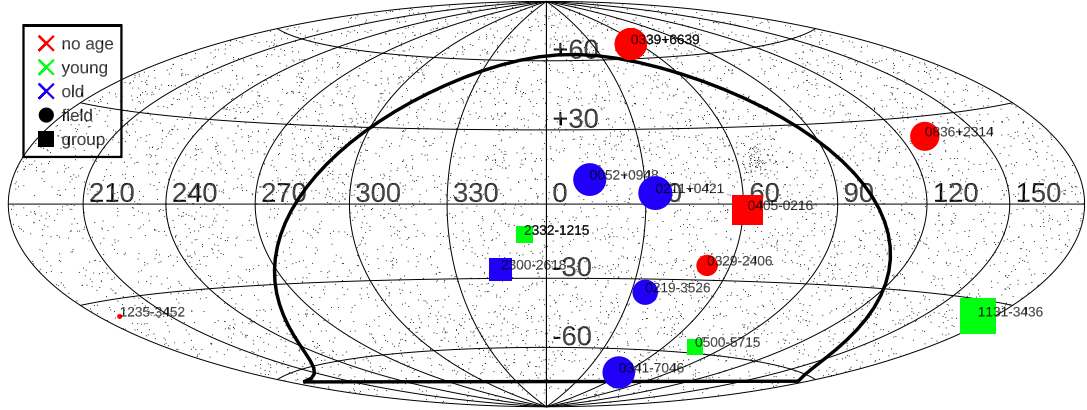}
\caption{An Aitoff projection displaying the sky positions for the 14 CYSs within 50 pc. Note that $0339+6639$ comprises two stars in a resolved binary system. The colour scheme provides an age estimate for the stars based on assessment of Li EW, where objects in red lack any Li EW measurement therefore have an indeterminate age, blue symbols are objects that have no indication of youth and green symbols are likely to be at least as young as the Pleiades. Square symbols indicate objects that belong to the potential MG member list in Table~\ref{table:MG_Candidates}, even if they are suggested to be non-members in \S\ref{sec:mg_candidates}. Circles represent stars that have no entry in Table~\ref{table:MG_Candidates} and are very likely to be nearby field stars. The relative sizes of the symbols represent distance, with the closest stars appearing the largest. The small black dots on the plot represent $\sim 6000$ K-stars within 50\,pc identified in DR1 and the enclosed solid-black loop is the Galactic Plane.}
\label{figure:Aitoff_50pc}
\end{figure*}

There are 14 CYS stars within 50\,pc. Almost all of these have been identified as members of multiple systems either from our duplicate RV analysis (see \S\ref{sec:binarity}), from identifying apparently equidistant, comoving DR2 companions or from the source identification in Simbad. An Aitoff projection for the CYS within 50\,pc is presented in Figure~\ref{figure:Aitoff_50pc}, in equatorial coordinates, where the symbol sizes represent the distances (largest symbol, $20.3\,$pc, $11315526-3436272$ and smallest symbol, 46.1\,pc, $12353357-3452547$) and small black filled circles represent the entire sample of K stars from the {\it Gaia} DR1 survey within 50\,pc. Of the 14 CYS within 50\,pc, 9 are not listed in Table~\ref{table:MG_Candidates} and have no connection to nearby MGs; these are denoted with circles. Objects represented by squares are potential MG members (present in Table~\ref{table:MG_Candidates}), even if they are subsequently rejected as members as a consequence of the analysis described in \S\ref{sec:mg_candidates}. Ten of the 14 CYS within 50\,pc sample are located in the Southern hemisphere, yet none are found in the right-ascension range $190^{\circ} < \alpha < 330^{\circ}$. This may reflect the avoidance of the Galactic plane in the GALEX survey. None of the 14 CYS within 50\,pc that are possible MG members are identified in the Northern hemisphere.

Of the 14 CYS within 50\,pc, only 8 stars have Li EW measurements from which ages might be inferred or constrained. Five of these 8 stars have Li EWs that suggest they are at least older than the Pleiades. Four of these five are likely field objects, while one, $23005681+3713528$, has a kinematic match with ABDMG, whose age is marginally older than the Pleiades; indeed, the Li EW of $23005681+3713528$ is consistent with that observed amongst similar ABDMG members. The other three stars have Li EWs that are consistent with the ages of clusters at least as young as the Pleiades. The Li EWs of these three stars match with stars of similar type in the MGs in which we have proposed membership ($11315526-3436272$ in TWA, $05004714-5715255$ and $23323085-1215513$ in BPMG; see $\S$\ref{sec:mg_candidates}).

Among the CYS within $50\,$pc, only 10 stars have counterparts in the RASS All-Sky-Survey, despite being presumably bright enough for detection. Of the stars with a RASS detection, only half have $\log f_{\rm X}/f_{\rm bol} > -4.0$, typical of K stars in young groups (see \S\ref{sec:xray}). Like the large $f_{\rm X}/f_{\rm bol}$~spread of the Table~\ref{table:Sample}~stars (\S\ref{sec:xray}), this suggests the CYSs include a significant number of apparently isochronally young stars ($< 80\,$Myr) that are UV-bright, but are weak X-ray emitters. We discuss the implications of this result in \S\ref{sec:discussion}.

\subsection{High probability young stars based on Li EW}\label{sec:Li_young_stars}

\begin{figure*}
\centering
\includegraphics[width=0.8\textwidth,angle=0]{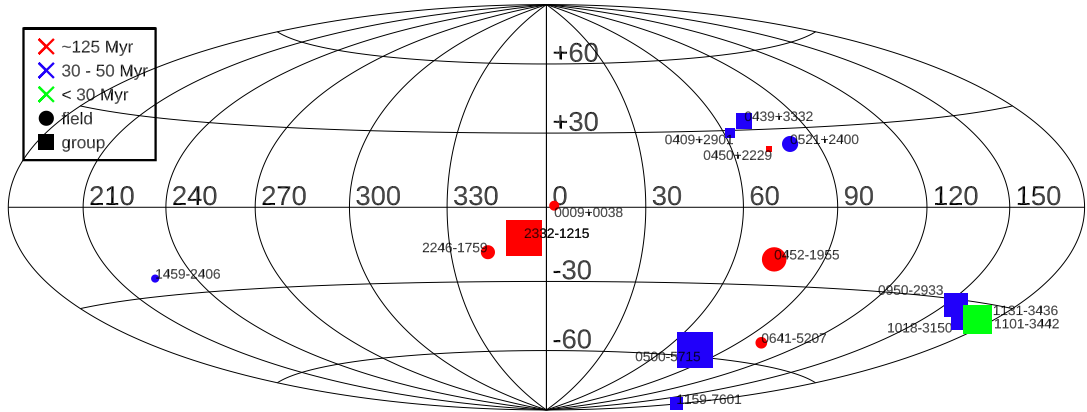}
\caption{An Aitoff projection displaying the sky positions for the 16 stars that have Li EW upper ages $\leq 125\,$Myr.}
\label{figure:Aitoff_Young}
\end{figure*}

In total, we identify 5 stars, at least as young as similar objects in the Pleiades, that are most likely to be young field stars. In \S\ref{sec:lithium} we identify 16 stars among the 146 CYS that have that have Li EW consistent with counterparts in the Pleiades or even younger groups. These are the objects in Table~\ref{table:Spectroscopy_Data} (final column) with Li EW ages $\leq 125\,$Myr. We plot all stars with upper Li EW ages $\leq 125\,$Myr on an Aitoff projection in Figure~\ref{figure:Aitoff_Young}. The sample is presented in a similar manner to Figure~\ref{figure:Aitoff_50pc} in terms of age, distance and group membership. There are 13 stars among these 16 that are present in Table~\ref{table:MG_Candidates} as potential MG members, although 2 of these, upon further assessment, $04524951-1955016$ and $09503676-2933278$ are more likely to be young field stars (\S\ref{sec:mg_candidates}). The 3 Li-rich stars that are not present in Table~\ref{table:MG_Candidates}, $00092179+0038065$, $06411248-5207385$ and $22464298-1759072$ are at least as young as 125\,Myr and potentially form part of the nearby young field star population. Both $00092179+0038065$ and $22464298-1759072$ are listed in the original SACY sample, but have no MG membership assigned in the literature. The star $06411248-5207385$ is suggested as a CAR member in \citet{2014a_Elliott}, but we find zero probability of MG membership predicted by BANYAN$\Sigma$ for any MG and neither in any of our kinematic tests.

From the 39 stars with an Li EW measurement, our 16 stars with Li-based ages at least as young as the Pleiades gives success rate in uncovering genuinely young stars of 40 per cent. It is possible that there remain a significant number of genuinely young field stars yet to be confirmed by their Li EW content. From the 6 stars for which we obtained spectra for the first time we detected one star with strong Li absorption. We plot all stars with upper Li EW ages $\leq 125\,$Myr on an Aitoff projection in Figure~\ref{figure:Aitoff_Young}. The sample is presented in a similar manner to Figure~\ref{figure:Aitoff_50pc} in terms of age, distance and group membership.

It is most likely that the majority of our CYS turn out to be unresolved binaries, and indeed, 30 per cent of the 146 CYS sample appear to be common proper-motion binaries (see \S\ref{sec:binarity}). Some of the earlier K-type stars match closely with a Pleiades binary sequence, and their multiplicity status could be identified by taking repeat spectroscopic observations to probe for variable radial velocity profiles.

\section{Discussion}\label{sec:discussion}

\subsection{New and previously established MG members: a surprisingly low yield}\label{sec:discussion_low_yield}

We have identified 146 CYSs in our UV- and {\it Gaia}-DR1-based search for nearby ($D \le 120\,$pc), isochronally-young (age $< 100\,$Myr), low-mass ($0.5 - 1.0\,M_{\odot}$) objects. The kinematics of our sample suggest that, despite their young isochronal ages, only $\sim 10$ per cent belong to nearby MGs. Upon closer analysis, we find that 16 stars are possible or probable members of young, nearby groups. As shown in Table~\ref{table:MG_Candidates}, out of our 23 candidate moving group members, two are members of the much older Hyades, and five stars are likely to be part of the field population. Three possible or probable MG members have no previous literature assignment to any known MG, and the membership status of each of these remains tenuous, awaiting confirmation via additional spectroscopic indicators of youth and/or RV measurement. Based on rigourous kinematic analysis and age-dating assessments, we identify one new candidate member of OCT, $21351099+3402313$, that shares similar $UVW$ and is located at a suitable distance, but requires stronger evidence of youth for confirmation and another new, but weaker OCT candidate, $22233510+3402313$, which requires both a RV measurement and further evidence of youth.

The fact that our CYS sample includes a handful of new MG candidates and more than a dozen previously identified nearby young stars, including the archetypical TW Hya, demonstrates the potential of our candidate selection criteria and followup kinematic test methods to identify such objects from among the field star population. Our yield of 16 young stars among 146 candidates (11 per cent) is comparable or better than those found in recent MG identification codes (\citealt{2014a_Gagne,2017a_Riedel,2019a_Bowler,2019a_Schneider}), where young star yields are generally far lower than 10 per cent. We note that there are 107 CYS that have no Li measurement as of yet, therefore our 11 per cent yield is a lower-limit, and much likely to be higher given that the hit-rate is $\sim 40$ per cent amongst the 39 stars with Li measurements. 

However, this yield still seems somewhat low, given that the sample of 146 was selected on the basis of {\it Gaia}-based isochronal ages $\leq 80\,$Myr (Figure~\ref{figure:K_GK_CMD}) as well as high levels of NUV excess (Figure~\ref{figure:NUV_Diff}). Indeed, Figures~\ref{figure:Xray_Clusters} and \ref{figure:LiEW_BP_RP} provide strong evidence that our sample of nearby young star candidates is in fact dominated by stars with ages older than the Pleiades, despite the fact that these stars lie high above the locus of MS stars in a {\it Gaia}/2MASS colour-magnitude diagram (Figure~\ref{figure:K_GK_CMD}). From the sample of Li-rich objects in this work, we find 5 that are likely to be isolated, young field stars. Such objects may be extremely useful to examine the processes by which young stars, predominantly born in clusters, are ejected from their nascent groups and dissolve into the field star population and could provide a fuller picture of the kine-dynamical nature of young stellar population in the Solar vicinity.

There are 159 objects in Figure~\ref{figure:G_BP_RP_CMD1} that were not included in our final CYS sample because either they had an insufficient number of visibility periods or failed the RUWE criteria (or both). Their positions on the DR2 CMD are not significantly different from the CYS sample, therefore we suspect that these other 159 objects are similarly `problematic', in terms of their combination of (over)luminosities and UV excesses (i.e. looking young) on the one hand and demonstrating weak X-rays and anomalous kinematics (i.e. looking old) on the other. We expect that including these objects in our CYS sample would not alter the outcomes found in this study.

\subsection{UV-bright and overluminous, but X-ray faint: a condundrum}

As noted in \S\ref{sec:xray}, the CYS sample of 146 stars appears to be dominated by stars with rather low coronal activity levels, based on their low overall RASS X-ray detection rate as well as the low X-ray fluxes of many of the individual stars that were detected. More specifically, the majority of the RASS-detected stars lie within $\sim 75$\,pc (Figure~\ref{figure:Xray_Clusters}), despite the fact that the RASS was capable of detecting K and early-M type stars of Pleiades or younger age out to $\sim 100$\,pc (e.g., \citealt{2013a_Rodriguez}, their figure 15 and table 6). The weak X-ray fluxes of a majority of the CYS sample is also reflected in the fact that 5 of the 14 stars within 50 pc have $\log{L_{\rm X}/L_{\rm bol}} < -4.0$, and another 4 were undetected in the RASS, suggesting even lower relative X-ray luminosities. That such a large fraction of the CYS sample display low coronal activity levels is surprising, given that all stars were selected on the basis of GALEX UV detections and young ($\le$80 Myr) isochronal ages. These results could have far-reaching implications for the use of {\it Gaia}-based isochronal ages to select CYSs for purposes of testing models of the manifestation and evolution of stellar magnetic activity, \citep[see, for example, figure 5 in][]{2018a_Zari}.

\begin{figure}
\centering
\includegraphics[width=0.45\textwidth,angle=0]{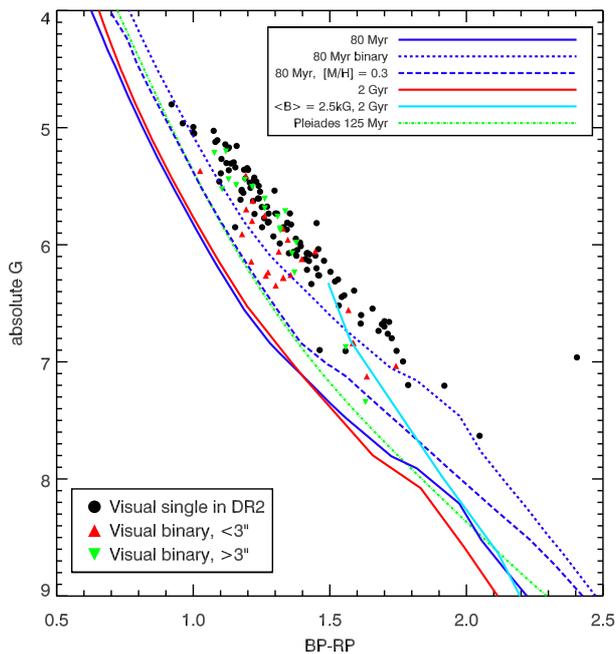}
\caption{The same absolute $G$ versus $G_{\rm BP} - G_{\rm RP}$ CMD as in Figure\,\ref{figure:G_BP_RP_CMD1} for the CYS sample, where visual single stars are displayed as black filled circles, close visual binaries as red upwards-facing triangles and wide visual binaries as green downwards-facing triangles (see $\S$\ref{sec:binarity}). The wide-dashed blue line represents a 80\,Myr with [Fe/H] = 0.3\,dex and the solid cyan line describes a 2\,Gyr isochrone with surface magnetic fields $<B> = 2.5$\,kG \citep[][see $\S$\ref{sec:discussion}]{2016a_Feiden}.}
\label{figure:G_BP_RP_CMD2}
\end{figure}

We can likely rule out the following five potential explanations for the non-young-star-like X-ray and kinematic properties of the majority of the candidates: 

\begin{enumerate} 
\item Contamination by first-ascent giant stars. This explanation would appear to be at odds both with the CMD distribution of the candidates (Figure~\ref{figure:K_GK_CMD}) and with the frequency of UV excesses among the candidate stars (Figure~\ref{figure:NUV_Diff}). {Our spectroscopic observations (\S\ref{sec:spectra}) also yielded no evidence of contamination of the CYS sample by giants.}
\item Binaries with a narrow range of separations. Binary stars with separations such that both stars are included in 2MASS photometry ($\sim 2''$ PSF), but only the primary is measured by {\it Gaia} at $G$ band, would shift an apparently single star upwards and to the right in Figure~\ref{figure:K_GK_CMD}. However, such confusion should only apply to a highly specific subsample of binary stars with separations around $\sim 1''$ which are likely to have been detected in DR2 (see $\S${sec:binarity}), and there appears to be very little effect on the CMD when plotted using purely DR2 photometry (Figure~\ref{figure:G_BP_RP_CMD2}).
\item Higher-order multiple systems. It would take {\it at least} an equal-mass quadruple system for a typical multiple MS system to appear even close to the locus of our CYSs. The rate of occurence for higher-order multiples ($N \geq 4$) is $\sim 3$\,per cent \citep[][]{2010a_Raghavan}, and these should have been resolved by DR2 in any case. Therefore, we rule out the higher-order multiple hypothesis.
\item MS stars with white dwarf (WD) companions. A subset of our candidates may be MS stars that have been spun up and/or inflated by accretion of mass lost by the asymptotic giant branch antecedent of a companion WD \citep{1996a_Jeffries}, such that the WD is in fact a contributor to (or dominates) the UV detected by GALEX. But it seems highly improbable that such systems would dominate our sample.
\item Super metal-rich ([Fe/H] $> 0.2$) stars. Although very rare, the most extreme known metal-rich stars have [Fe/H] $\sim 0.3$\,dex \citep[][]{2001a_Feltzing}, and stars with higher metallicities have larger luminosities. In Figure\,\ref{figure:G_BP_RP_CMD2} we plot a 80\,Myr (PARSEC 3.1) isochrone at this metallicity; it is apparent that the CYS sample stars are much brighter than this isochrone. A combination of both high metallicity and multiplicity may in some cases place MS stars close to the locus of our CYS sample. However, it is exceedingly unlikely that a significant number of our CYSs are in this category.
\end{enumerate} 

In the absence of a better explanation, the surprising ``bifurcation'' of our UV- and isochronally-selected nearby young star candidates into X-ray-bright and X-ray-faint subsamples (\S\ref{sec:xray}; Figure~\ref{figure:Xray_Clusters}) raises a set of particularly vexing questions. Namely: If the large (apparently dominant) fraction of X-ray-faint stars among our candidates are in fact not pre-MS stars, then why are they as ``overluminous'' (in terms of their bolometric luminosities) as the X-ray-bright stars?

Could (at least some of) our sample be ``imposters'' -- zero-age (or even older) MS stars that are ``puffed up'' via high levels of magnetic activity due to fast rotation rates, as appears to be the case for, e.g., the subset of overluminous late-type Pleiades stars \citep{2017a_Somers}? We compare our CYS sample with a $2\,$Gyr single-star isochrone from the Dartmouth evolutionary models, with strong surface magnetic fields \citep[][$<B> = 2.5$\,kG]{2016a_Feiden}, since strong magnetic fields are expected to inflate radii of K/M-type stars thus increasing their bolometric luminosities \citep[][]{2017a_Somers, 2018a_Jackson}\footnote{The outputs from the magnetic models give $\log L$ and $\log T_{\rm eff}$, so we use the EEM table to obtain the appropriate $G_{\rm BP}-G_{\rm RP}$ and $M_{\rm G}$ values}. The magnetic models only go as massive as $M/M_{\odot} = 0.8$, however we can still identify that the low-mass stars (early K-type and beyond) generally lie above the magnetic MS, however some of the sample lie within the bounds of the magnetic models. It is within reason that MS stars with large magnetic fields could partially explain overluminosity (with $<B> = 2.5$\,kG, see Figure\,\ref{figure:G_BP_RP_CMD2}), however older stars are highly unlikely to be so strongly magnetically active \citep[for example,][find no early MS Solar-type stars with surface magnetic fields $> 0.2$\,kG]{2018a_Folsom}, and in any case, if they have strong magnetic fields they should have been supported by evidence of activity, e.g., strong H$\alpha$ emission, which we do not observe for the vast majority of our CYS sample.

It is furthermore possible that these apparently overluminous stars manifest significant photometric variability. If this variability were at levels of a few tenths of a magnitude, they might transiently appear more luminous than their non-variable (older) counterparts. There are 80 sources in our CYS that have SuperWASP identifiers within 10'' and more than 1000 photometric data points. All but 6 of these have both standard deviations and 5-95$^{\rm th}$ percentile ranges less than 0.2\,mag, and the other 6 appear to have large photometric errors. Even if these stars did demonstrate high levels of photometric variability, why do so many of our magnetically active, presumably radially inflated young MS field stars have such weak coronae relative to ``normal'' young stars, despite their apparently similar levels of chromospheric UV excess?

Could some or all of these stars be rotationally inflated yet X-ray faint due to centrifugally induced ``coronal stripping'' \citep{2004a_Jardine}? This is highly unlikely as it would apply to ultra-fast rotating stars that are super-saturated in X-ray. These are only slightly less X-ray active than the ``saturated'' X-ray emitters ($\log L_{\rm X}/L_{\rm bol} \sim -3$), and would have periods of $\leq 0.5$\,day (and would still be very spotted, and very active). Such stars would have stood out in X-ray surveys and many would have been previously found as active stars with measured rotation periods due to spottedness. Such characteristics are notably lacking for most of the CYS sample.

Addressing these questions will require a dedicated observing campaign targeting our candidate stars in the optical through UV to X-ray regimes, so as to access diagnostics of chromospheric and coronal activity and relate these properties to stellar age indicators and rotation rates. Ultraviolet spectroscopy with HST, as well as Chandra and XMM X-ray spectroscopy, represent the additional, essential puzzle pieces necessary to understand the natures of the class of isochronally young and UV-bright yet X-ray faint and kinematically ``old'' stars uncovered by this work. Regardless, the work presented here, like that of \citet[][]{2018a_Wright}, hints at the power of {\it Gaia} astrometric and photometric data for purposes of isolating the population of young MS stars that originated in young moving groups and have relatively recently mixed into the field star population.

\section{Summary and Conclusions}\label{sec:summary}

We have identified a kinematically and almost spatially unbiased sample of 146 nearby, isochronally-young, UV-excessive K-type stars, and we conduct astrometric and age-dating analyses to identify their astrophysical nature. We find only $\sim 10$ per cent of the sample can be classed as highly-likely young stars and in most cases one must find alternative scenarios to explain their overluminosity compared to typical MS counterparts. A small, but presumably representative spectroscopic sample uncovered just one new young star. This spectroscopic sample provided a useful supplement for measuring Galactic kinematics.

About three-quarters of the full sample of 146 stars have sufficient data to measure Galactic space velocities. We identify 16 of these stars that satisfy kinematic criteria for membership to at least one nearby MG. Five are new candidate members with no previous identification of MG membership. There are 5 stars with Li EW measurements consistent with age upper-limits of $\sim 125\,$Myr which appear to be isolated, young, nearby field-stars.

Puzzlingly, a significant fraction of the UV-bright, overluminous stars here identified as CYSs either either do not have X-ray detections or have very low X-ray activity, despite being close enough ($D \leq 50\,$pc) to surpass X-ray sensitivity limits. These objects, whether young or more evolved, appear to pose significant challenges to our present understanding of magnetic activity in late-type stars. Additional UV and X-ray spectroscopic observations are required to ascertain the natures of these overluminous, UV-bright, yet X-ray-weak stars.

Finally, we reiterate (see \S\ref{sec:selection}) that the same selection criteria used to isolate the 146 stars in Table~\ref{table:Sample}~from {\it Gaia} DR1 data would yield in excess of 1500 candidate stars, if applied to DR2 data. In addition to yielding a larger number of new candidate MG members, investigation of this DR2 ``supersample'' would likely yield greater insight into the puzzling UV versus X-ray behaviors of overluminous field stars.

\section*{Acknowledgments}

ASB acknowledges the financial support of the STFC and UNAM. This research was supported by NASA Astrophysics Data Analysis Program (ADAP) grant NNX12AH37G to RIT and UCLA and NASA ADAP grant NNX09AC96G to RIT, and by a National Science Foundation Research Experience for Undergraduates program grant to RIT's Center for Imaging Science that supported M. Chalifour's summer 2017 RIT residency. The authors thank R. Jeffries for helpful discussions. The authors thank the anonymous referees for useful suggestions and comments, which have significantly improved this manuscript.

\bibliographystyle{mnras}
\bibliography{bibliography}

\end{document}

%% file: T_Sample.tex
{\tiny
\begin{table*}
  \caption{The 2MASS identifier names, spectral types based on linear interpolation of $G-K_{\rm s}$ colour (SpT$_{\rm P}$), spectral types given in SIMBAD (SpT$_{\rm S}$), {\it Galex} NUV magnitudes, DR2 photometry, 2MASS\,K magnitudes, X-ray to bolometric luminosities and UV excesses for the candidate young stars. The first 10 stars are listed here and the remaining stars can be accessed in the electronic version of this publication.}
\begin{center}
\begin{tabular}{lrrrrrrrrr}
  \hline
  \hline
Name               & SpT$_{\rm P}$ & SpT$_{\rm S}$ & $NUV$              & $G$       & $G_{\rm BP}$ & $G_{\rm RP}$ & $K_{\rm s}$        & $\log(f_x/f_{\rm bol})$ & $\Delta UV$ \\
(2MASS J-)         &           &           & (mag)              & (mag)     & (mag)        & (mag)        & (mag)              &                         & (mag)       \\
  \hline
$00044817-4959504$ & K3.1      &           & $18.992 \pm 0.056$ & $10.8804$ &    $11.4445$ &    $10.1998$ &  $8.768 \pm 0.021$ &                         &     $+0.04$ \\
$00092179+0038065$ & K4.0      & K4Ve      & $18.290 \pm 0.052$ & $11.0253$ &    $11.6454$ &    $10.3281$ &  $8.711 \pm 0.019$ &                 $-3.21$ &     $-1.54$ \\
$00475278-3245205$ & K3.7      & K2Ve      & $16.696 \pm 0.024$ & $10.1600$ &    $10.7268$ &     $9.4748$ &  $7.899 \pm 0.029$ &                 $-3.10$ &     $-5.79$ \\
$00524693+0948123$ & K1.6      & K0        & $14.659 \pm 0.008$ &  $8.3148$ &     $8.7536$ &     $7.7503$ &  $6.476 \pm 0.027$ &                 $-4.77$ &     $-0.76$ \\
$01011333-4517578$ & K3.2      & K0V       & $15.674 \pm 0.012$ &  $9.6079$ &     $9.9796$ &     $8.9547$ &  $7.501 \pm 0.026$ &                         &     $-2.04$ \\
$01445006-0805455$ & K1.3      &           & $17.260 \pm 0.025$ & $10.5147$ &    $10.9449$ &     $9.9450$ &  $8.703 \pm 0.019$ &                         &     $-0.34$ \\
$01513522+0827126$ & K2.6      &           & $18.318 \pm 0.055$ & $10.5858$ &    $11.1220$ &     $9.9358$ &  $8.574 \pm 0.021$ &                         &     $+0.12$ \\
$01524398-7445462$ & K3.1      & K3V       & $16.531 \pm 0.019$ &  $9.7562$ &    $10.3018$ &     $9.1044$ &  $7.675 \pm 0.018$ &                 $-3.58$ &     $-1.14$ \\
$02115797+0421416$ & K4.7      & K5        & $17.236 \pm 0.017$ &  $9.4273$ &    $10.0699$ &     $8.7081$ &  $7.035 \pm 0.026$ &                 $-3.61$ &     $-1.25$ \\
$02194778-3526443$ & K5.5      & K2+Vk     & $15.109 \pm 0.008$ &  $8.6211$ &     $9.0810$ &     $7.9023$ &  $6.084 \pm 0.021$ &                 $-5.02$ &     $-3.26$ \\
\hline
\end{tabular}
\end{center}
\label{table:Sample}
\end{table*}}

%% file: T_Spectroscopy_Observations.tex
{\centering
\begin{table*}
\caption{Subsample observed spectroscopically. Objects labelled with asterisks do not form part of the CYS sample because they did not satisfy the Gaia DR2 astrometric criteria. Spectral types are measured using the {\sc sptclass} software \citep[][]{2017a_Hernandez}. SNR values for MIKE spectra are the interquartile range. Where no Li EW was observed the 2$\sigma$ upper limit is given.}
\begin{center}
\begin{tabular}{lrrrrrrrr}
\hline
\hline
                 Name &       $G$ &       $d$ &          HJD &  SpT &                   RV &      SNR &        Li EW & H$\alpha$ EW \\
           (2MASS J-) &     (mag) &      (pc) &   (2450000-) &      & (${\rm km\,s}^{-1}$) &          &       (m\AA) &        (\AA) \\
\hline
\multicolumn{9}{c}{San Pedro de Mart\'ir 2.1-m Telescope -- Echelle Spectrograph} \\
$0016+3031^{\rm *}$   &  $8.5088$ &  $39.276$ & $7934.96117$ & K4.3 &     $+2.02 \pm 0.14$ & $22$     &       $< 56$ &      $+0.84$ \\
$0052+0948$           &  $8.3148$ &  $44.982$ & $7936.95071$ & K0.2 &                      & $16$     &       $< 78$ &      $+1.00$ \\
$0211+0421$           &  $9.4273$ &  $46.825$ & $7937.96785$ & K4.5 &    $-65.36 \pm 4.65$ & $20$     &       $< 62$ &      $+0.08$ \\
\hline
\multicolumn{9}{c}{Australian National University Telescope -- WiFeS} \\
$0052+0948$           &  $8.3148$ &  $44.982$ & $8364.11052$ & G9.0 &       $-0.9 \pm 0.7$ & $47$     &       $< 26$ &      $+1.63$ \\
$0211+0421$           &  $9.4273$ &  $46.825$ & $8363.25773$ & K3.5 &      $-57.3 \pm 0.7$ & $30$     &       $< 42$ &      $+0.11$ \\
$0219-3526^{\rm a}$   &  $8.6211$ &  $34.870$ & $8363.27311$ & K1.1 &       $+2.1 \pm 0.7$ & $35$     &       $< 36$ &      $+1.43$ \\
$0235+1514^{\rm *}$   &  $9.2371$ &  $43.422$ & $8363.27869$ & K3.5 &      $-29.5 \pm 0.6$ & $27$     &       $< 46$ &      $+0.76$ \\
                      &           &           &              &      &      $-76.8 \pm 1.0$ & $39$     &       $< 32$ &              \\
$0341-7046$           & $10.5656$ &  $44.101$ & $8363.30671$ & K9.2 &       $+1.5 \pm 0.7$ & $23$     &       $< 54$ &      $+0.57$ \\
$0341+0336^{\rm *}$   &  $8.9882$ &  $25.833$ & $8363.24749$ & K2.6 &       $+0.4 \pm 0.8$ & $21$     &       $< 60$ &      $+0.56$ \\
$0341-5125^{\rm b,*}$ &  $8.6087$ &  $36.112$ & $8363.31628$ & K3.5 &      $+51.3 \pm 0.7$ & $39$     &       $< 32$ &      $+0.86$ \\
$2014-3936^{\rm c,*}$ & $11.2289$ &  $91.501$ & $8364.12422$ & K6.0 &       $-9.5 \pm 1.0$ & $30$     &       $< 42$ &      $+0.74$ \\
$2059-4758^{\rm *}$   &  $9.9843$ &  $42.495$ & $8364.04859$ & K7.3 &       $+3.9 \pm 0.7$ & $23$     &       $< 54$ &      $+0.59$ \\
$2142-3035^{\rm *}$   &  $9.9022$ &  $31.835$ & $8364.05803$ & K5.5 &      $+14.0 \pm 0.6$ & $24$     &       $< 52$ &      $+0.60$ \\
$2251-4646^{\rm d,*}$ &  $9.1318$ &  $42.524$ & $8364.07792$ & K1.0 &      $-34.3 \pm 0.7$ & $37$     &       $< 34$ &      $+1.09$ \\
$2332-1215^{\rm e}$   &  $9.8182$ &  $27.373$ & $8364.11616$ & M0.6 &       $+3.0 \pm 0.8$ & $17$     & $152 \pm 18$ &      $-1.83$ \\
\hline
\multicolumn{9}{c}{Magellan Telescope -- MIKE Echelle Spectrograph} \\
$0950-2933^{\rm f}$   & $11.0389$ & $121.068$ & $8567.79963$ & K2.3 &    $+14.10 \pm 1.36$ & $34-80$  &  $197 \pm 4$ &      $-0.24$ \\ 
$1018-3150$           & $10.8465$ &  $65.644$ & $8567.81306$ & K8.7 &    $+15.74 \pm 4.57$ & $33-103$ &  $473 \pm 4$ &      $-3.36$ \\ 
$1037-0623^{\rm *}$   &  $9.4856$ &  $37.866$ & $8567.80740$ & K9.1 &     $-9.43 \pm 0.24$ & $36-104$ &        $< 7$ &      $+0.71$ \\ 
$1131-3436$           & $10.4296$ &  $49.379$ & $8567.79963$ & M0.4 &     $+8.04 \pm 1.47$ & $33-173$ &  $547 \pm 3$ &     $-14.89$ \\ 
$1159-7601$           & $10.8095$ &  $99.747$ & $8567.89020$ & K5.2 &    $+14.86 \pm 0.30$ & $41-111$ &  $429 \pm 3$ &      $-0.21$ \\ 
$1235-3452^{\rm f}$   &  $7.4906$ &  $21.681$ & $8567.83803$ & K1.9 &    $-37.81 \pm 0.27$ & $72-170$ &        $< 4$ &      $+0.87$ \\ 
$1306-4609^{\rm *,b}$ & $11.6020$ &  $98.899$ & $8567.88194$ & K3.2 &    $+12.42 \pm 2.41$ & $23-74$  &       $< 11$ &      $-2.78$ \\ 
$1459-2406$           & $11.2699$ & $113.817$ & $8567.91591$ & K4.9 &     $-1.24 \pm 0.31$ & $38-94$  &  $314 \pm 4$ &      $+0.08$ \\ 
$1637+2919^{\rm f,g}$ & $11.0190$ & $101.130$ & $8567.79963$ & K5.2 &    $-32.31 \pm 0.74$ & $32-102$ &        $< 7$ &      $-0.66$ \\ 
                      &           &           &              &      &    $-48.82 \pm 1.59$ &          &              &              \\
$1935-1502^{\rm *}$   & $11.0877$ & $104.858$ & $8567.89839$ & K6.5 &     $-5.73 \pm 0.37$ & $44-121$ &        $< 6$ &      $+0.72$ \\ 
$2032-4742^{\rm *}$   & $10.0732$ &  $31.265$ & $8567.91042$ & K8.0 &     $-9.43 \pm 0.18$ & $36-116$ &        $< 7$ &      $+0.62$ \\ 
\hline
\end{tabular}
\begin{flushleft}	
a = VB, unresolved; b = Broad CCF, minimal asymmetry, SB?; c = nearby neighbour to the north west; d = VB, extracted both components (A = North); e = Known member of BPMG; f = double-lined CCF; g = RV resolved into two components. In this Table, and those that follow, for brevity we use only the first four characters in the right ascension and declination of 2MASS names; full designations are listed in column 1 of Table~\ref{table:Sample}.
\end{flushleft}
\label{table:Spectroscopy_Observations}
\end{center}
\end{table*}}

%% file: T_Spectroscopy_Data.tex
{\centering
\begin{table}
  \caption{H$\alpha$ and Li EW data.}
\begin{center}
\begin{tabular}{p{1.55cm}p{0.9cm}p{0.7cm}p{0.2cm}p{0.65cm}p{0.2cm}p{1.15cm}}
\hline
\hline
Name        & ${\rm BP-RP}$ & H$\alpha$EW &  ref &   LiEW &  ref & LiEW age   \\
(2MASS J-)  & (mag)         & (\AA)       &      & (m\AA) &      & (Myr) \\
\hline
$0009+0038$ &      $1.3173$ &      $-0.2$ &    a &  $100$ &    a & $\sim 125$ \\
$0047-3245$ &      $1.2520$ &      $-2.0$ &    a &    $0$ &    a &            \\
$0052+0948$ &      $1.0033$ &    $+1.315$ &   *: &  $<52$ &   *: &            \\
$0152-7445$ &      $1.1974$ &             &      &    $0$ &    a &            \\
$0211+0421$ &      $1.3618$ &    $+0.095$ &   *: &   $52$ &   *: &            \\
$0219-3526$ &      $1.1787$ &     $+1.43$ &    : &  $<36$ &    : &            \\
$0339+6639$ &      $1.3296$ &    $+0.376$ &    b &   $64$ &    b &            \\
$0341-7046$ &      $1.6304$ &     $+0.57$ &    : &  $<54$ &    : &            \\
$0409+2901$ &      $1.1195$ &     $+0.95$ &  cde &  $345$ &    c &    $10-30$ \\
$0423+2940$ &      $1.2618$ &    $-0.063$ &   df &        &      &            \\
$0439+3332$ &      $1.5023$ &      $-0.7$ &  dgh &  $495$ &    i &    $10-30$ \\
$0443-4106$ &      $1.1032$ &      $-0.2$ &    a & $< 50$ &    a &  $125-625$ \\
$0450+2229$ &      $1.1730$ &      $+1.4$ &    e &  $275$ &    i &   $20-125$ \\
$0452-1955$ &      $1.3066$ &             &      &  $190$ &    j & $\sim 125$ \\
$0500-5715$ &      $1.9194$ &    $-1.178$ &  akl &  $340$ &  ano &    $20-50$ \\
            &               &             &    m &        &    p &            \\
$0521+2400$ &      $1.2219$ &     $-1.07$ &   dq &  $380$ &   dq &    $10-30$ \\
$0641-5207$ &      $1.1771$ &      $+0.0$ &    a &  $180$ &    a & $\sim 125$ \\
$0815+2946$ &      $1.1777$ &    $-0.158$ &    r &   $20$ &    r &            \\
$0833+3350$ &      $1.1589$ &    $+0.392$ &    r &   $33$ &    r &            \\
$0950-2933$ &      $1.2178$ &     $-0.22$ &  a\$ &  $249$ &  a\$ &   $20-125$ \\
$0959+3849$ &      $1.2310$ &    $-0.694$ &    r & $< 17$ &    r &            \\
$1018-3150$ &      $1.7120$ &     $-3.32$ &  aks &  $517$ &  ano &    $10-30$ \\
            &               &             &  t\$ &        &   \$ &            \\
$1101-3132$ &      $1.1925$ &      $+0.0$ &    a &    $0$ &    a &            \\
$1101-3442$ &      $1.6563$ &      $+166$ &  akm &  $441$ &  ano &    $10-30$ \\
            &               &             &  suv &        &    s &            \\
$1131-3436$ &      $2.4045$ &     $-8.29$ &  akm &  $553$ &  ano &     $< 20$ \\
            &               &             &  suv &        & sw\$ &            \\
            &               &             & wx\$ &        &      &            \\
$1133+3613$ &      $1.4580$ &    $-0.672$ &    r &   $10$ &    r &            \\
$1159-7601$ &      $1.4515$ &    $-0.417$ &  agu &  $445$ &  ajy &    $10-30$ \\
            &               &             &  y\$ &        &   \$ &            \\
$1215-0237$ &      $1.7867$ &    $-1.037$ &  kmz &        &      &            \\
$1221+2005$ &      $1.2243$ &    $+0.842$ &    r &    $8$ &    r &            \\
$1224-7503$ &      $1.3908$ &      $-0.4$ &    a &    $0$ &    a &            \\
$1235-3452$ &      $1.2734$ &     $+0.87$ &   \$ &  $< 7$ &   \$ &            \\
$1459-2406$ &      $1.3168$ &     $+0.08$ &   \$ &  $314$ &   \$ &    $10-30$ \\
$1637+2919$ &      $1.4156$ &    $-0.456$ &  r\$ &  $< 7$ &  r\$ &            \\
$1757+5844$ &      $1.3975$ &     $-0.11$ &    r & $< 11$ &    r &            \\
$1916-5328$ &      $1.3442$ &      $+0.0$ &    a &    $0$ &    a &            \\
$2154+2239$ &      $1.7442$ &     $-0.70$ &    m &        &      &            \\
$2223+3231$ &      $1.3776$ &    $+0.134$ &    f &        &      &            \\
$2246-1759$ &      $1.1834$ &      $+0.0$ &    a &  $160$ &    a & $\sim 125$ \\
$2252-6843$ &      $1.2510$ &      $+0.0$ &    a &   $50$ &    a &  $125-625$ \\
$2259-4900$ &      $1.3777$ &      $-0.3$ &    a &    $0$ &    a &            \\
$2300-2618$ &      $1.7422$ &      $-0.1$ &    u &   $15$ &   ao &            \\
$2330-1717$ &      $1.1294$ &      $+0.0$ &    a &   $30$ &    a &  $125-625$ \\
$2332-1215$ &      $2.0483$ &     $-1.66$ &  akm &  $169$ &   a: &   $20-125$ \\
            &               &             &   u: &        &      &            \\
\hline
\end{tabular}
\begin{flushleft}	
References: a = \citet{2006a_Torres}; b =  \citet{2018a_Frasca}; c = \citet{2007a_White}; d = \citet{2017a_Kraus}; e = \citet{1996a_Wichmann}; f = \citet{2016a_Luo}; g = \citet{2010a_Wahhaj}; h = \citet{2015a_Alonso-Floriano}; i = \citet{2000a_Wichmann}; j = \citet{2009a_daSilva}; k = \citet{2015a_Ansdell}; l = \citet{2017a_Zerjal}; m = \citet{2014a_Gaidos}; n = \citet{2008a_Fernandez}; o = \citet{2008a_Mentuch}; p = \citet{2010a_Yee}; q = \citet{1998a_Li}; r = \citet{2015a_Binks}; s = \citet{2012a_Schneider}; t = \citet{2017a_Fang}; u = \citet{2006a_Riaz}; v = \citet{2013a_Dent}; w = \citet{2017a_Riedel}; x = \citet{2018a_Riedel}; y = \citet{2007a_Guenther}; z = \citet{2012a_Schlieder}; * = SPM; : = ANU; \$ = MIKE.
\end{flushleft}
\label{table:Spectroscopy_Data}
\end{center}
\end{table}}

%% file: T_CPMB.tex
{\tiny
\begin{table*}
  \caption{Stars with probable common proper motion companions in DR2, where column 2 describes the angular separation between the components. The first 10 stars are listed here and the remaining stars can be accessed in the electronic version of this publication.}
\begin{center}
\begin{tabular}{lrrrrr}
  \hline
  \hline
Name                & r      & $\pi$                & $\mu_{\alpha}$       & $\mu_{\delta}$       & $G$                 \\
(2MASS J-)          & (as)   & (mas)                & (mas/yr)             & (mas/yr)             & (mag)               \\
  \hline
$0009+0038$        & $15.6$ &  $9.2716 \pm 0.0828$ & $+121.978 \pm 0.120$ &  $-27.048 \pm 0.097$ & $11.0253 \pm 0.0038$ \\
                   &        &  $9.4633 \pm 0.6243$ & $+125.936 \pm 1.099$ &  $-24.733 \pm 0.711$ & $19.1824 \pm 0.0044$ \\
$0101-4517$        &  $0.5$ & $14.1960 \pm 0.0269$ & $+129.511 \pm 0.034$ &  $+26.077 \pm 0.036$ &  $9.6079 \pm 0.0003$ \\
                   &        & $14.0202 \pm 0.0420$ & $+123.878 \pm 0.062$ &  $+22.986 \pm 0.049$ & $10.4047 \pm 0.0009$ \\
$0211+0421$        & $20.3$ & $21.3561 \pm 0.0993$ & $-144.316 \pm 0.121$ &  $-78.640 \pm 0.154$ &  $9.4273 \pm 0.0004$ \\
                   &        & $21.3649 \pm 0.0519$ & $-144.381 \pm 0.075$ &  $-78.861 \pm 0.072$ & $12.6154 \pm 0.0003$ \\
$0219-3526$        &  $1.9$ & $28.6784 \pm 0.0355$ &  $-88.820 \pm 0.030$ &  $+39.343 \pm 0.049$ &  $8.6211 \pm 0.0003$ \\
                   &        & $28.6758 \pm 0.0394$ &  $-90.351 \pm 0.036$ &  $+17.384 \pm 0.061$ &  $8.8098 \pm 0.0003$ \\
$0241-3735$        &  $0.3$ & $11.7480 \pm 0.0262$ &  $+27.857 \pm 0.030$ &  $-42.562 \pm 0.042$ & $11.7740 \pm 0.0005$ \\
                   &        & $11.7587 \pm 0.0359$ &  $+33.984 \pm 0.039$ &  $-44.060 \pm 0.081$ & $12.4029 \pm 0.0005$ \\
$0339+6639{\rm a}$ &  $1.9$ & $22.8417 \pm 0.0312$ &  $-79.635 \pm 0.027$ &  $-21.683 \pm 0.048$ &  $9.4879 \pm 0.0004$ \\
                   &        & $22.8510 \pm 0.0344$ &  $-74.190 \pm 0.030$ &   $-8.488 \pm 0.051$ &  $9.8160 \pm 0.0005$ \\
$0339+6639{\rm b}$ &  $1.9$ & $22.8510 \pm 0.0344$ &  $-74.190 \pm 0.030$ &   $-8.488 \pm 0.051$ &  $9.8160 \pm 0.0005$ \\
                   &        & $22.8417 \pm 0.0312$ &  $-79.635 \pm 0.027$ &  $-21.683 \pm 0.048$ &  $9.4879 \pm 0.0004$ \\
$0341-4516$        &  $1.2$ &  $8.0760 \pm 0.0218$ &  $+23.824 \pm 0.034$ &  $-25.397 \pm 0.042$ & $11.5209 \pm 0.0010$ \\
                   &        &  $6.6154 \pm 0.3144$ &  $+22.447 \pm 0.672$ &  $-26.569 \pm 0.688$ & $15.8465 \pm 0.0092$ \\
$0341-7046$        &  $9.7$ & $22.6749 \pm 0.0285$ &  $+78.573 \pm 0.050$ & $+145.128 \pm 0.058$ & $10.5656 \pm 0.0007$ \\
                   &        & $22.6765 \pm 0.0437$ &  $+73.354 \pm 0.083$ & $+154.921 \pm 0.088$ & $14.7510 \pm 0.0006$ \\
$0409+2901$        &  $6.8$ &  $9.0651 \pm 0.0415$ &  $+23.800 \pm 0.093$ &  $-34.904 \pm 0.046$ & $10.4169 \pm 0.0013$ \\
                   &        &  $9.1769 \pm 0.0435$ &  $+24.635 \pm 0.093$ &  $-34.090 \pm 0.050$ & $13.5890 \pm 0.0009$ \\
\hline
\end{tabular}
\end{center}
\label{table:CPMB}
\end{table*}
}

%% file: T_Kinematics.tex
{\tiny
\begin{table*}
\caption{Kinematic data for the candidate young stars. The precision of the measurements are given to one decimal place, however, are generally more precise. The first 10 stars are listed here and the remaining stars can be accessed in the electronic version of this publication, where their full precision values are quoted.}
\begin{center}
\begin{tabular}{lrrrrrrrr}
  \hline
  \hline
Name       & ${\mu}_{\alpha}$       & ${\mu}_{\delta}$       & RV                   & flags       & $\pi$          & $U$                  & $V$                  & $W$                  \\
(2MASS J-) & (${\rm mas\,yr}^{-1}$) & (${\rm mas\,yr}^{-1}$) & (${\rm km\,s}^{-1}$) &             & (mas)          & (${\rm km\,s}^{-1}$) & (${\rm km\,s}^{-1}$) & (${\rm km\,s}^{-1}$) \\
  \hline
$0004-4959$ & $+190.6 \pm 0.3$      &  $-24.3 \pm 0.6$       & $-55.2 \pm 1.3$      & a,1         & $8.7 \pm 0.3$  & $-104.9 \pm 0.7$     & $-44.6 \pm 0.4$      & $+36.2 \pm 1.2$      \\
$0009+0038$ & $+124.1 \pm 3.5$      &  $-26.8 \pm 1.5$       & $-31.9 \pm 5.8$      & a,1         & $8.9 \pm 0.9$  & $-44.6 \pm 0.8$      & $-54.7 \pm 2.8$      & $+11.1 \pm 5.1$      \\
$0047-3245$ & $+246.9 \pm 0.2$      &  $-11.9 \pm 0.1$       & $+1.0 \pm 3.0$       & b,1         & $13.7 \pm 0.3$ & $-69.5 \pm 0.3$      & $-49.3 \pm 0.2$      & $-1.8 \pm 1.0$       \\
$0052+0948$ &  $-33.4 \pm 0.1$      &  $-49.4 \pm 0.1$       & $-0.9 \pm 0.7$       & :,1,**      & $22.6 \pm 0.3$ & $+10.9 \pm 0.2$      & $-3.6 \pm 0.4$       & $-5.7 \pm 0.6$       \\
$0101-4517$ & $+129.6 \pm 0.1$      &  $+25.7 \pm 0.1$       & $-15.3 \pm 0.4$      & b,1,**      & $14.6 \pm 0.3$ & $-41.1 \pm 0.1$      & $-13.1 \pm 0.1$      & $+13.7 \pm 0.4$      \\
$0144-0805$ &  $-37.2 \pm 2.6$      &  $-50.0 \pm 1.0$       & $+22.907 \pm 1.035$  & ac,3        & $8.1 \pm 0.6$  & $+27.1 \pm 0.4$      & $-2.4 \pm 0.2$       & $-35.9 \pm 1.0$      \\
$0151+0827$ &  $+38.6 \pm 2.1$      &   $+9.4 \pm 0.6$       & $+36.85 \pm 0.36$    & c,1         & $9.0 \pm 0.3$  & $-36.2 \pm 0.2$      & $+3.5 \pm 0.1$       & $-21.3 \pm 0.3$      \\
$0152-7445$ &  $+61.9 \pm 1.1$      &  $+83.5 \pm 0.8$       & $+6.778 \pm 2.853$   & a,1,WUMa    & $13.8 \pm 0.2$ & $-31.5 \pm 1.0$      & $-5.4 \pm 1.9$       & $-21.2 \pm 1.9$      \\
$0211+0421$ & $-144.4 \pm 0.1$      &  $-78.9 \pm 0.1$       & $-57.479 \pm 6.296$  & *:,5        & $21.7 \pm 0.2$ & $+62.3 \pm 3.5$      & $-5.3 \pm 1.4$       & $+27.0 \pm 5.0$      \\
$0219-3526$ &  $-87.0 \pm 0.1$      &  $+39.9 \pm 0.1$       & $+1.593 \pm 0.500$   & b:,3        & $28.1 \pm 0.3$ & $+5.2 \pm 0.1$       & $+13.5 \pm 0.2$      & $-7.1 \pm 0.5$       \\
\hline
\end{tabular}
\begin{flushleft}
RV references: a = \citet{2017a_Kunder}; b = \citet{2006a_Gontcharov}; c = \citet{2018a_Gaia_Collaboration}; d = \citet{2018a_Frasca}; e = \citet{2012a_Nguyen}; f = \citet{1994a_Barbier-Brossat}; g =  \citet{2006a_Malaroda}; h = \citet{2006a_Torres}; i = \citet{2014a_Elliott}; j = \citet{2001a_Montes}; k = \citet{2013a_Murphy}; l = \citet{1995a_Duflot}; m = \citet{2010a_Schlieder}; n = \citet{2016a_Sperauskas}; o = \citet{2014a_Malo}; * = measured from spectrum collected at the SPM; : = measured from spectrum collected at the ANU. Binary flags: 0 = no RV measurements, binary status is indeterminate; 1 = 1 RV measurement, binary status is indeterminate; 3 = $> 2$ RV measurements, all of which are within $5\,{\rm km\,s}^{-1}$ of one another and their errors (added in quadrature) are $< 5\,{\rm km\,s}^{-1}$ and are likely single stars; 5 = $> 2$ RV measurements, of which at least one pair are separated by $> 5\,{\rm km\,s}^{-1}$ of one another and their errors (added in quadrature) are $< 5\,{\rm km\,s}^{-1}$ and are likely to be binary stars; 9 = errors are too large to determine their binary status. Simbad flags are: ** = double star; BYDra = BY Draconis; WUMa = eclipsing binary star of type W Ursa Majoris; EB* = eclipsing binary; SB* = spectroscopic binary star; RSCVn = RS Canum Venaticorum variable star; Algol = eclipsing binary of Algol type; $\beta$Lyr = eclipsing binary of $\beta$ Lyrae type.
\end{flushleft}
\label{table:Kinematics}
\end{center}
\end{table*}}

%% file: T_MG_Candidates.tex
{\centering
\begin{table*}
  \caption{Candidate MG members.}
\begin{center}
\begin{tabular}{lrrrrrrrrrr}
\hline
\hline
Name        & Age            & Distance  & kinematic  &    $t_{1}$ &  $t_{2}$ &  $t_{3}$ & $t_{4}$ & ${\rm MG_{\rm lit}}$ & BANYAN & Final \\
(2MASS J-)  & (Myr)          & (pc)      & match(es)  &            & (pc)     & (${\rm km\,s}^{-1}$) & &             &      & choice   \\
\hline
\multicolumn{10}{c}{Stars with a measured RV and pass all kinematic criteria.} \\
\hline 
$0409+2901$ &    $10-30^{*}$ & $110.313$ &        THO &     $1.22$ &  $-0.25$ &  $-0.87$ &  $0.54$ &                  & TAU(13.7) & TAU \\
            &                &           &        COL &     $3.60$ &  $-8.02$ &  $+1.46$ &  $0.76$ &                  &           &     \\
$0439+3332$ &    $10-30^{*}$ &  $90.092$ &        THO &     $0.69$ &  $-0.20$ &  $-3.18$ &  $0.18$ &                  &           & TAU \\
            &                &           &        EPS &     $0.47$ &  $-1.42$ &  $-0.38$ &  $0.67$ &                  &           &     \\
            &                &           &        TWA &     $2.39$ &  $+7.79$ &  $-1.69$ &  $0.76$ &                  &           &     \\
$0500-5715$ &    $20-50^{*}$ &  $26.901$ &        BPM &     $0.03$ &  $+0.75$ &  $+0.08$ &  $0.09$ & BPM$^{\rm abcd}$ & BPM(99.9) & BPM \\
            &                &           &        TWA &     $2.78$ &  $-4.91$ &  $+0.59$ &  $0.60$ &                  &           &     \\
$0521+2400$ &                &  $88.267$ &        THO &     $0.58$ &  $-0.28$ &  $-2.39$ &  $0.24$ &                  & 118(99.8) & 118 \\
            &                &           &        EPS &     $0.44$ &  $-4.40$ &  $+0.39$ &  $0.78$ &                  &           &     \\
            &                &           &        TWA &     $3.04$ &  $+7.70$ &  $-0.75$ &  $0.72$ &                  &           &     \\
$0950-2933$ &   $20-125^{*}$ & $121.068$ &        THO &     $1.53$ &  $-5.84$ &  $-2.30$ &  $0.91$ &                  &           & YFS \\
$1018-3150$ &    $10-30^{*}$ &  $65.644$ &        COL &     $1.15$ &  $+4.58$ &  $+0.96$ &  $0.29$ &   TWA$^{\rm ab}$ &           & TWA \\
            &                &           &        TWA &     $1.53$ &  $+8.38$ &  $+4.10$ &  $0.23$ &                  &           & /COL \\
$1101-3442$ &    $10-30^{*}$ &  $60.086$ &        COL &     $3.24$ &  $+0.72$ &  $-2.78$ &  $0.23$ & TWA$^{\rm abcd}$ & TWA(99.9) & TWA \\
            &                &           &        TWA &     $0.79$ &  $+5.18$ &  $+0.23$ &  $0.12$ &                  &           &     \\
$1121-4736$ &      $< 10-40$ &  $70.723$ &        THO &     $1.26$ &  $+9.09$ &  $+1.15$ &  $0.86$ &                  &           & IFS \\
            &                &           &        ABD &     $2.42$ &  $+0.55$ &  $-7.56$ &  $0.58$ &                  &           &     \\
$1131-3436$ &         $< 20$ &  $49.379$ &        BPM &     $2.51$ &  $+1.03$ &  $+0.48$ &  $0.32$ & TWA$^{\rm abcd}$ &           & TWA \\
            &                &           &        TWA &     $1.21$ &  $+2.45$ &  $-2.48$ &  $0.12$ &                  &           &     \\
$1159-7601$ &    $10-30^{*}$ &  $99.747$ &        THO &     $1.67$ &  $-9.20$ &  $+2.01$ &  $0.16$ &    CAR$^{\rm d}$ & EPS(99.9) & EPS \\
            &                &           &        EPS &     $0.02$ &  $+1.32$ &  $+0.02$ &  $0.11$ &                  &           &     \\
$1215-0237$ &        $10-60$ &  $51.590$ &        ABD &     $2.76$ &  $+3.41$ &  $+2.42$ &  $0.39$ &                  & ABD(19.7) & ABD \\
$2135+3402$ &        $15-60$ &  $75.764$ &        OCT &     $0.96$ &  $+0.76$ &  $-1.84$ &  $0.15$ &                  &           & OCT \\
$2332-1215$ &   $20-125^{*}$ &  $27.373$ &        BPM &     $0.41$ &  $-0.49$ &  $+0.94$ &  $0.10$ &  BPM$^{\rm abc}$ & BPM(99.9) & BPM \\
\hline
\multicolumn{10}{c}{One star without a measured RV, with $\chi^{2} < 3.78$ for {\it some} RV configuration and pass all other kinematic criteria.} \\
\hline
$2223+3231$ &                &  $85.497$ &        OCT &   8.5,12.5 &  $-9.25$ &          &  $0.11$ &                  &           & OCT \\
\hline
\multicolumn{10}{c}{Stars classed as MG members in literature, missed in our search} \\
\hline
$0405-0216$ &        $25-80$ &  $40.289$ &            &            &          &          &  $0.62$ &                  & HYA(48.9) & HYA \\
$0418+2317$ &        $10-60$ &  $54.008$ &            &            &          &          &  $0.12$ &                  & HYA(96.5) & HYA \\
$0450+2229$ &   $20-125^{*}$ & $123.684$ &            &            &          &          &  $0.11$ &                  & TAU(99.8) & TAU \\
$0452-1955$ & $\sim 125^{*}$ &  $61.686$ & THA(74.23) &            & $-14.39$ & $+43.77$ &  $0.29$ &    THA$^{\rm c}$ &           & YFS \\
$1459-2406$ &        $10-30$ & $113.817$ &            &            &          &          &  $0.13$ &                  & UCL(94.6) & UCL \\
            &                &           &            &            &          &          &  $0.29$ &                  &  USC(0.3) &     \\
$2213+8445$ &                &  $61.523$ &            &            &  $+0.27$ &          &  $0.37$ &                  & ARG(66.3) & IFS \\
$2300-2618$ &        $10-60$ &  $31.859$ &  ABD(6.63) &            &  $+1.75$ &  $+2.56$ &  $0.19$ &                  & ABD(96.7) & ABD \\
\hline
\end{tabular}
\begin{flushleft}	
All candidate young stars that have some kinematic match to a known MG, either from our analyses, has a candidate or member status from a literature source, or both. Columns $t_{1-4}$ are the results of the kinematic criteria described in \S\ref{sec:kinematic_tests}. Top section: objects with at least one RV measurement and satisfy all kinematic criteria given in \S\ref{sec:kinematic_tests}. Middle section: objects that lack any RV measurement but would have some RV value satisfying the $\chi^{2}$-test. We provide the RV range corresponding to a $\chi^{2}$-test success in the $\chi^{2}$ column. Bottom section: objects that fail the $\chi^{2}$ test, however are identified as MG members either as candidate MG members in the literature and/or with probabilities $> 10$ per cent for a given MG based on the BANYAN\,$\Sigma$ code. If the potential host MG suggested in the literature is one of the 11 MGs analysed in this work, then the $\chi^{2}$ value (always $> 3.78$) is provided in the $\chi^{2}$ column. The abbreviations used for the groups in this work are as follows: ABD = AB Doradus, ARG = Argus, BPM = $\beta$ Pictoris, COL = Columba, EPS = $\epsilon$~Cha, ETA = $\eta$~Cha, HYA = Hyades, LCC = Lower Centaurus Crux, OCT = Octans, TAU = Taurus, THA = Tucana Horologium, THO = 32 Ori, TWA = TW Hya, UCL = Upper Centaurus Lupus{\bf , USC = Upper Scorpius}. References: a = \citet{2013a_Malo}; b = \citet{2015a_Bell}; c = \citet{2004a_Zuckerman}; d = \citet{2008a_Torres}. Ages with superscripted asterisks are those measured using Li EWs (see Table~\ref{table:Spectroscopy_Data} and Figure~\ref{figure:LiEW_BP_RP} for reference). Final designations YFS and IFS are "young field stars" and "indeterminate-age field star", respectively. Note that $1018-3150$ is also a possible member of COL.
\end{flushleft}
\label{table:MG_Candidates}
\end{center}
\end{table*}}